# Techno-Economic Analysis of Repurposing Abandoned Oil Wells for Geothermal Energy Extraction Using Physics-Informed Neural Networks

Hung-Yu Lin[1], Kuan-Chun Shih[2], Lea-Der Chen[1]

[1]Texas A&M University – Corpus Christi, Corpus Christi, Texas, USA
[2]National Taiwan University, Taipei, Taiwan

**Conference note:** Presented at the ASME 2026 20th International Conference on Energy Sustainability (ES2026), Bellevue, Washington, USA, July 26–29, 2026. Paper No. ES2026-184639.

## ABSTRACT

*To achieve net-zero targets by 2050, it is critical to diversify renewable energy. Hydropower, wind, and solar energy dominate; geothermal energy remains underutilized. Conventional Enhanced Geothermal Systems (EGS) rely on hydraulic stimulation, which poses risks such as induced seismicity. To address this, Closed-Loop Geothermal Systems (CLGS) circulate working fluids in sealed tubing to avoid direct reservoir contact, making them a potential solution for repurposing idle oil wells without environmental hazards. This study developed a Physics-Informed Neural Network (PINN) to model the CLGS performance. Unlike traditional neural networks, PINN explicitly embeds governing physical equations into their learning processes, such as heat conduction and convection. This integration enabled the model to accurately predict the wellbore temperature and flow characteristics over a 25-year lifespan, even with sparse training data. The simulation results confirmed stable long-term predictions. When coupled with an Organic Rankine Cycle (ORC) model, the system yielded a thermodynamic efficiency of 9.5%. Crucially, several economic indicators (e.g., DPP, NPV, and LCOE) are conducted to evaluate the investment feasibility and economic potential of the proposed CLGS-based power generation system. This proposed framework provides a scalable, physics-consistent tool for evaluating both technical performance and economic returns, offering a robust pathway to accelerate geothermal adoption.*

## 1. INTRODUCTION

### 1.1 Background research & motivation

The rapid advancement of digital technologies and the proliferation of energy-intensive applications have led to a significant increase in global electricity demand. According to the Energy Information Administration (EIA), electricity consumption in the United States is projected to continue its upward trend in 2024–2025, driven not only by traditional residential and commercial usage but also by the exponential growth of artificial intelligence workloads and cryptocurrency mining operations [1]. Simultaneously, the urgency to decarbonize power generation has led to a notable increase in renewable energy integration. The share of renewables in U.S. electricity generation is expected to grow from 21% in 2023 to 24% in 2024 (EIA, 2024) [2]. with geothermal energy emerging as a promising candidate owing to its base-load stability, low greenhouse gas emissions, and independence from weather variability. A 2024 study by the U.S. The Department of Energy (DOE) has projected that geothermal electricity capacity could reach 90 GW by 2050 under favorable policy and technological conditions [3], showing the substantial impact that further development of geothermal resources could have on the future energy system.

### 1.2 Challenges in geothermal energy deployment

Despite the long-term stability of geothermal energy, its development in deep or low-permeability formations faces significant challenges, complicating site identification. Conventional Enhanced Geothermal Systems (EGS) rely on natural or stimulated fracture networks for fluid circulation. However, uncertainties in permeability, fracture connectivity, and flow pathways lead to variability in thermal output and system performance, thereby increasing exploration and operational risks [4, 5]. These uncertainties often require deeper drilling and reservoir stimulation, thereby further increasing capital costs and financial risk before productivity can be confirmed [6].

Therefore, closed-loop geothermal systems (CLGS) have been proposed as an alternative configuration in which working fluids circulate within sealed wellbores, reducing dependence on in-situ reservoir permeability [7]. By decoupling heat extraction from formation fluid flow, CLGS mitigate risks associated with uncertain hydraulic properties and expand feasible deployment sites [8]. Its structural similarity to conventional oil and gas wells also enables repurposing of abandoned wells, reducing capital costs by avoiding new drilling [9, 10].

CLGS, however, remains relatively immature, with limited large-scale and long-term validation [7]. As a result, data for model calibration are scarce, limiting purely data-driven machine learning approaches that require extensive datasets [11]. Most existing studies rely on numerical methods, such as FEM [12] and FVM [13], to simulate coupled flow and heat transfer. While accurate, these methods are computationally expensive for long-term prediction and multi-scenario analysis. They do not provide readily available models for rapid evaluation, limiting their applicability in early-stage design and feasibility assessment [13, 14].

These limitations highlight the need for a computationally efficient, physics-informed predictive framework for CLGS analysis under limited data conditions, also enabling techno-economic evaluation of system feasibility. A Physics-Informed Multi-Model Framework for CLGS Assessment is presented in this paper.

To address these challenges, a physics-informed multi-model framework is developed to assess the repurposing of abandoned oil wells into a coaxial closed-loop geothermal system, with the objective of providing a computationally efficient, economically oriented evaluation tool.

A transient one-dimensional wellbore heat transfer model [15] is first formulated based on energy conservation principles to capture the dominant thermal processes, including axial convection of the working fluid and radial heat conduction into the surrounding formation. This model is used to generate synthetic data across a range of geological and operational conditions, forming the basis for training a Physics-Informed Neural Network (PINN).

Unlike conventional artificial neural networks, which depend primarily on large datasets, the PINN incorporates governing physical equations directly into the training process through automatic differentiation. This allows the model to maintain physical consistency while significantly reducing data requirements. Compared to traditional numerical simulations, the PINN serves as a surrogate model that preserves the underlying physics while enabling rapid predictions of long-term system behavior with substantially lower computational cost [11, 16, 17, 18].

To convert the extracted geothermal heat into usable electrical power, an Organic Rankine Cycle (ORC) model is adopted. ORC systems are particularly well suited for low- to medium-temperature heat sources [12, 20]. They use organic working fluids with lower boiling points than water, enabling efficient energy conversion under moderate thermal conditions. The closed-loop geothermal systems, for example, those repurposed from abandoned oil and gas wells, operate at lower outlet temperatures than conventional high-temperature geothermal reservoirs. The ORC provides a practical solution for power generation in CLGS applications.

Finally, an economic model is developed to evaluate the financial feasibility and cost competitiveness of the proposed CLGS scenarios. Discounted cash flow–based indicators, including Net Present Value (NPV) and Discounted Payback Period (DPP) [21], are used to calculate capital intensity and long project lifetimes [20]. In addition, the Levelized Cost of Electricity (LCOE) [21, 22] is calculated to assess the average cost of electricity generation over the system lifetime and enable comparison with other energy technologies [23].

By integrating physics-informed machine learning, thermal modeling, and techno-economic evaluation, this paper provides a systematic approach for the design and performance assessment of CLGS. The proposed methodology enhances predictive reliability, supports long-term thermal performance analysis, and enables quantitative evaluation of economic feasibility, thereby facilitating informed engineering decisions prior to system deployment.

## 2. METHODOLOGY

### 2.1 Overview of the integrated modeling framework

The framework depicted in The figure 1 illustrates the workflow of each cycle within the integrated modeling framework, highlighting the interactions and data flow among the three main models. The PINN model predicts the outlet temperature of the geothermal fluid after heat exchange, based on input parameters such as rock formation, wellbore, working fluid, and operating time. Utilizing a physics-informed neural network, the PINN ensures physically consistent and stable temperature predictions. The ORC model takes the predicted hot water temperature from the PINN as input and calculates the system's electricity generation, accounting for thermodynamic efficiencies and environmental factors to estimate power output. The economic analysis model integrates the ORC power generation results with development and maintenance costs derived from PINN parameters, such as well depth, casing material, and operational expenses. This model determines the optimal geothermal working temperature that minimizes the payback period, thereby optimizing economic feasibility. The entire process forms a closed-loop cycle, in which cooled water from the ORC model is reinjected into the PINN model (well) to initiate the next prediction cycle, enabling continuous system performance and economic optimization.

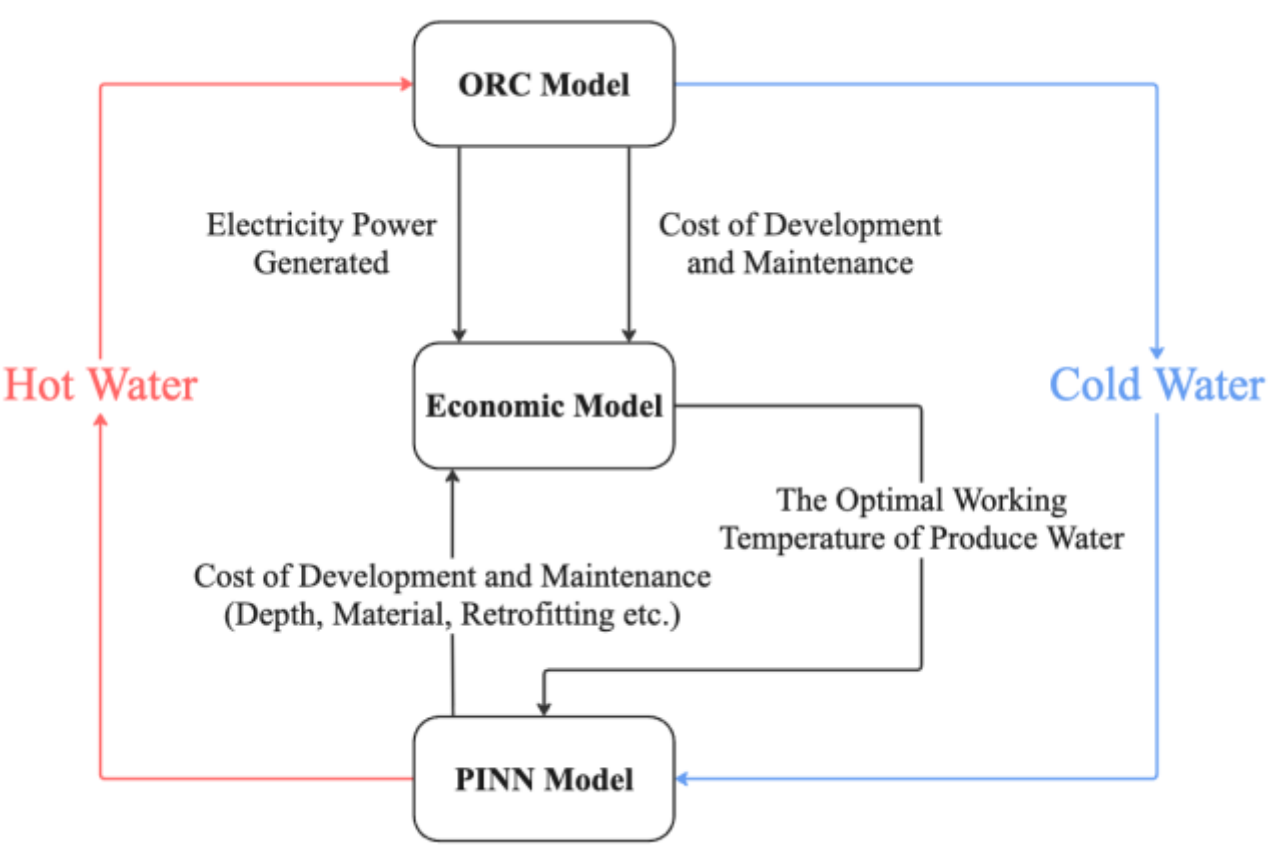

The figure 1. Overall framework and workflow of the coupled ORC, Economic, and PINN models

### 2.2 SYNTHETIC DATA GENERATION

Due to the limited availability of historical field data for CLGS, a simulation program based on a transient one-dimensional wellbore heat-transfer model [15] is used to generate data across various operational scenarios. The generated datasets are used to support PINN model training and validation.

The closed-loop geothermal system in this study, as shown in Figure 2, adopts a coaxial double-pipe wellbore configuration in which the working fluid circulates within a sealed system. Water is injected from the surface, flows downward through the annulus, receives geothermal heat, and returns through the inner pipe, forming a closed-loop circulation. The system configuration is based on previously reported CLGS designs [24, 25].

The wellbore is divided into sections based on the temperature difference between the circulating fluid and the surrounding formation. In the shallow section, where the formation temperature is lower than the injected fluid temperature, a cement layer is applied to provide thermal insulation and reduce heat loss. This insulation extends to the depth at which the formation temperature first exceeds the injection temperature. At greater depths, the cement layer is omitted to allow direct heat transfer between the steel casing and the surrounding formation, which serves as the primary heat-exchange region. The casing extends the full well depth and serves as the primary conductive interface. As the heated fluid returns to the surface through the inner pipe, heat exchange with the annulus may occur. To minimize this effect, an insulation layer is installed between the inner pipe and annulus to reduce internal heat transfer and prevent thermal short-circuiting.

The system is modeled as a closed-loop circulation process. To account for long-term thermal depletion, the formation temperature is updated annually

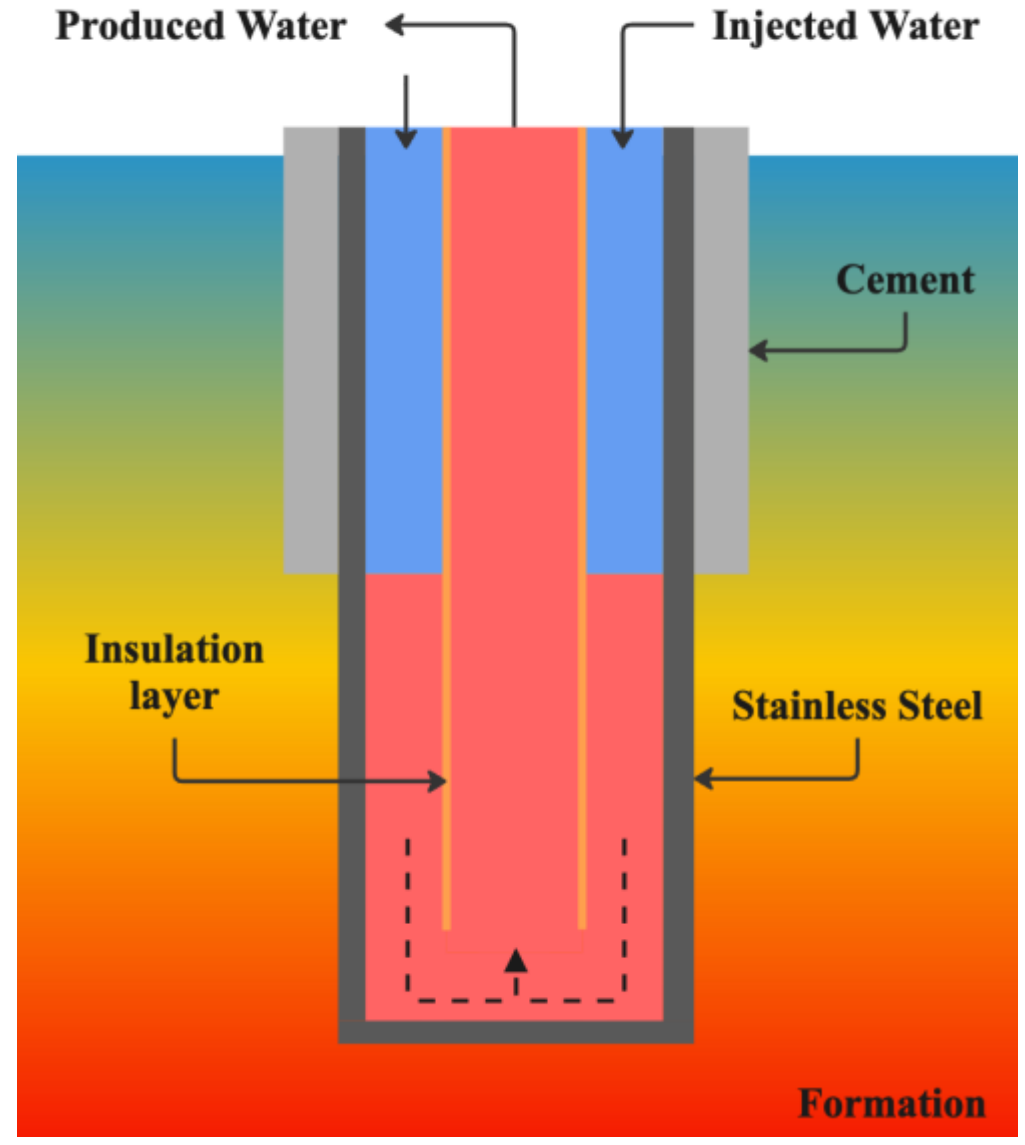

Figure 2. Schematic of the coaxial closed-loop geothermal well configuration

To generate physically consistent subsurface boundary conditions, a geothermal gradient range of 50–150 °C km⁻¹ was selected based on the central distribution of historical Icelandic borehole data from the Orkustofnun database [26]. While the temperature distribution with depth is data-informed, the thermal properties of the surrounding formation are specified using representative values from the literature for Icelandic volcanic rocks. The adopted thermal conductivity (1.5–3.0 W $m^{-1}$ $K^{-1}$) falls within reported ranges for basaltic rocks [27], and the corresponding thermal diffusivity ($0.7–1.5 \times 10^{-6}$ $m^2$ $s^{-1}$) is consistent with values derived from Icelandic rock property data [28].

Heat transfer along the wellbore is evaluated. Radial heat transfer between the annular fluid and the surrounding formation is modeled using a series thermal resistance model [29]. The total thermal resistance consists of contributions from the steel casing, cement sheath, and surrounding rock:

$$R_{total} = R_{steel} + R_{cement} + R_{rock} \tag{1}$$

| Symbol | Description | Unit |
|---|---|---|
| $R_{steel}$ | Thermal resistance of steel casing | K $W^{-1}$ |
| $R_{cement}$ | Thermal resistance of cement casing | K $W^{-1}$ |
| $R_{rock}$ | Thermal resistance of surrounding rock formation | K $W^{-1}$ |

The effective segment-wise heat transfer coefficient is given by:

$$UA_{rock} = \frac{\Delta z}{R_{total}} \tag{2}$$

| Symbol | Description | Unit |
|---|---|---|
| $\Delta z$ | Axial length of a discretized well segment | m |

The annulus temperature after heat exchange with the formation is updated as:

$$T^*_{annulus} = T_{rock,i} + (T_{annulus,i} - T_{rock,i})exp\left(-\frac{UA_{rock}}{\dot{m}c_p}\right) \tag{3}$$

| Symbol | Description | Unit |
|---|---|---|
| $T_{rock,i}$ | Formation (rock) temperature at segment $i$ | °C |
| $T_{annulus,i}$ | Annulus fluid temperature at segment $i$ | °C |
| $\dot{m}$ | Mass flow rate of the circulating fluid | kg $s^{-1}$ |
| $c_p$ | Specific heat capacity of the working fluid | J $kg^{-1}$ $K^{-1}$ |

After accounting for heat exchange between the annular fluid and the surrounding formation, heat transfer between the inner pipe and the annulus is subsequently considered. The inner pipe diameter is selected such that its cross-sectional area is equal to that of the annulus. Under this condition, both flow paths have the same fluid velocity for a given mass flow rate. Heat exchange between the two flow paths is modeled using the effectiveness–NTU method [30]. The segment-wise heat transfer coefficient across the insulation layer is given by:

$$\mathrm{UA_{pipe}} = \frac{\Delta \mathrm{z}}{\mathrm{R_{ins}}} \tag{4}$$

| Symbol | Description | Unit |
|---|---|---|
| $R_{ins}$ | Thermal resistance of insulation layer between inner tube and annulus | K W-1 |

Since both flow paths have the same cross-sectional area, their mass flow rates are identical, resulting in $C_r = 1$. Under this condition, the effectiveness of the heat exchanger simplifies to:

$$\varepsilon = \frac{\mathrm{NTU}}{1+\mathrm{NTU}}, \quad NTU = \frac{UA_{pipe}}{\dot{m}c_p} \tag{5}$$

where NTU is the number of transfer units, defined as the ratio of the overall heat transfer conductance to the thermal capacity rate.

At each segment, the heat exchange between the two flow paths is computed using the effectiveness relation. The temperature updates are given by:

$$T_{tube,i-1} = T_{tube,i} - \varepsilon(T_{tube,i} - T_{annulus,i}) \tag{6}$$
$$T_{annulus,i+1} = T_{annulus,i} + \varepsilon(T_{tube,i} - T_{annulus,i}) \tag{7}$$

| Symbol | Description | Unit |
|---|---|---|
| $T_{tube,i}$ | Inner fluid temperature at segment $i$ | ℃ |

where the inner pipe temperature is updated in the upward direction, while the annulus temperature is updated in the downward direction, consistent with the counterflow configuration.

The temperatures of both flow paths are updated sequentially along the well depth, subject to the bottom boundary condition, where $N$ denotes the deepest axial segment of the discretized wellbore:

$$T_{tube,N} = T_{annulus,N} \tag{8}$$

The heat exchange processes described above determine the thermal interaction between the circulating fluid and the surrounding formation. To capture the long-term cooling of the reservoir, the formation temperature is updated at each simulation year. Heat is continuously transferred from the formation to the circulating fluid, leading to gradual thermal depletion of the reservoir.

The heat extracted from the formation at each segment is evaluated based on the temperature difference between the rock and the annular fluid, together with the local thermal resistance:

$$q_i = \frac{T_{rock,i} - T_{annulus,i}}{R_{total,i}} \tag{9}$$

| Symbol | Description | Unit |
|---|---|---|
| $q_i$ | Heat transferred from rock to fluid in segment $i$ | W |
| $R_{total,i}$ | Total thermal resistance in segment $i$ | K W-1 |

The corresponding annual energy extracted from each segment is then calculated by integrating the heat flux over the segment length and time:

$$E_i = q_i' \Delta z \Delta t \tag{10}$$

| Symbol | Description | Unit |
|---|---|---|
| $E_i$ | Extracted thermal energy from rock in segment $i$ | J |
| $\Delta t$ | Time step used for annual energy calculation | s |
| $q_i^i$ | Heat transfer rate per unit length | W m-1 |

This extracted energy results in a reduction in the local rock temperature. The temperature change is determined from the thermal capacity of the rock:

$$\Delta T_{rock,i} = \frac{E_i}{V_i \rho_{rock} c_{rock}} \tag{11}$$

| Symbol | Description | Unit |
|---|---|---|
| $\Delta T_{rock,i}$ | Temperature reduction of rock in segment $i$ | ℃ |
| $E_i$ | Extracted thermal energy from rock in segment $i$ | J |
| $V_i$ | Cylindrical rock control volume associated with segment $i$ | $m^3$ |
| $\rho_{rock}$ | Rock density | $kg\ m^{-3}$ |
| $c_{rock}$ | Rock specific heat capacity | J kg-1 K-1 |
| $T_{surface}$ | Surface temperature | ℃ |

The updated rock temperature is obtained by subtracting the temperature drop from the previous value, with a lower bound imposed by the surface temperature:

$$T_{rock,i}^{new} = max(T_{surface}, T_{rock,i} - \Delta T_{rock,i}) \tag{12}$$

| Symbol | Description | Unit |
|---|---|---|
| $T_{surface}$ | Surface temperature | ℃ |

This formulation captures the progressive cooling of the surrounding formation over time while maintaining numerical stability in long-term simulations.

Since the heat transfer processes and rock temperature update are defined, the numerical solution procedure is implemented. The wellbore is discretized into $N$ axial segments. The number of segments is determined from the well depth $z$ to ensure sufficient spatial resolution while limiting computational cost:

$$N = min\left(max\left(\frac{z}{\Delta z}, N_{min}\right), N_{max}\right) \quad (13)$$

| Symbol | Description |
|---|---|
| $N_{min}$ | The minimum grid number |
| $N_{max}$ | The maximum grid number |

For each operational year, the solution is advanced sequentially along the wellbore. The calculation consists of three main steps: annulus-to-formation heat exchange, annulus-to-inner-pipe heat exchange, and annual updating of the surrounding rock temperature to account for thermal depletion.

The resulting temperature evolution profiles are used to generate physics-consistent synthetic datasets for training the Physics-Informed Neural Network (PINN), thereby enabling efficient approximation of long-term wellbore thermal behavior under sparse-data conditions.

## 2.3 PINN MODEL

Physics-Informed Neural Networks (PINNs) are used to model the thermal behavior of the closed-loop geothermal system by embedding governing physical laws into the learning process [11]. The input and output variables used for PINN training are summarized in, with each input variable selected to represent key physical and operational factors governing the system's behavior.

The input and output variables used for PINN training are summarized in Table 1, where each input variable is selected to represent key physical and operational factors that govern the system's behavior. The well depth determines the length of the heat-exchange pathway and directly controls the circulating fluid's exposure to subsurface geothermal gradients, thereby exerting a primary influence on the achievable outlet temperature. The outer annulus wellhead diameter characterizes the geometric configuration of the coaxial well and affects the available cross-sectional area for fluid circulation, which in turn influences flow velocity, convective heat transfer, and pressure losses along the wellbore.

The injection flowrate represents the operational condition of the system and governs the balance between residence time and convective heat transfer, playing a critical role in determining the temporal evolution of fluid temperature. The injection temperature specifies the thermal state of the working fluid entering the CLGS system and serves as a boundary condition for heat transfer between the fluid and the surrounding formation. Operating time is included to capture the transient thermal response of the closed-loop system, accounting for the progressive heat extraction from the formation and the resulting temporal variation in system temperatures. The segmented geothermal gradient is introduced to represent depth-dependent variations in subsurface temperature, enabling the model to capture non-uniform temperature distributions along the wellbore. The thermal conductivity and thermal diffusivity of the surrounding formation are included to characterize heat transfer within the rock, thereby governing both the rate of heat conduction toward the wellbore and the formation's temporal response to thermal extraction.

Table 1. Input and output variables used for PINN training

| Variable | Symbol | Unit | Description |
|---|---|---|---|
| Input Features | | | |
| Well depth | Z | m | Total vertical depth of the well |
| Injection flow rate | $\dot{m}$ | kg $s^{-1}$ | Mass flow rate of injected fluid |
| Injection (Inlet water) temperature | $T_{in}$ | °C | Temperature of injected fluid |
| Operating time | t | year | Elapsed operating time |
| Inner tube diameter | $D_{inner}$ | m | Diameter of the inner production tube. |
| Rock thermal conductivity | $k_{rock}$ | W $m^{-1}$ K | Thermal conductivity of the surrounding rock formation |

| Variable | Symbol | Unit | Description |
|---|---|---|---|
| Rock thermal diffusivity | $\alpha_{rock}$ | $m^2\ s^{-1}$ | Thermal diffusivity of the surrounding rock formation |
| Segmented geothermal gradients (0–10 km) | g= [g0…, g9] | °C $km^{-1}$ | Geothermal gradient in each 1-km depth interval from 0–10 km |
| | | Output Labels | |
| Wellhead annulus temperature | $T_{outer_head}$ | °C | Inlet temperature at the wellhead |
| Wellhead inner temperature | $T_{inner_head}$ | °C | Outlet temperature at the wellhead |

The model outputs are limited to the wellhead temperatures of the inner (tube) and outer (annulus) pipes. This design choice is intentional, as it allows the PINN to learn the system's underlying physical behavior without being explicitly provided with temperature information along the wellbore. By restricting the supervised outputs to wellhead measurements only, the model must infer the internal temperature evolution using the embedded physics-based constraints during training. Although the injection temperature and the outer wellhead temperature may take identical values at the surface, they represent different physical meanings in the modeling framework. The injection temperature corresponds to a known boundary condition at the wellhead, which can be directly specified from operational settings. In contrast, the outer wellhead temperature represents an observable system response at the surface. As the working fluid travels along the wellbore, its temperature evolution at depth is not prescribed to the model but is determined by the governing physical equations. Accordingly, during training, the total well depth is discretized into multiple segments, and the PINN predicts the temperature distribution along the wellbore at these internal locations to satisfy the physics constraints. This strategy enables the reconstruction of the full temperature profile while maintaining supervision only at the wellhead, thereby providing a rigorous assessment of the PINN's ability to capture the coupled thermal processes of the coaxial CLGS.

The network is implemented as a fully connected feedforward neural network (FNN). Let the input vector be denoted as $x = (z, p)$ where z represents the local axial coordinate along the wellbore (i.e., the current depth position) and p denotes the input of operational and geological parameters.

To account for spatial variability, collocation point $z$ is randomly sampled along the wellbore depth, allowing the model to enforce physical constraints and learn temperature distributions throughout the domain. In the temporal dimension, short-term fluid transport effects within the wellbore are relatively minor, and the dominant behavior is governed by long-term thermal depletion of the surrounding formation. Time is introduced as an input variable representing the operational duration, enabling the model to capture the long-term evolution of system temperatures.

The variable $z$ describes the spatial location along the well, while $Z$ denotes the total well depth, treated as a global parameter included in p. Accordingly, $z \in [0, Z]$, where $Z$ remains constant for a given scenario. The segmented geothermal gradient vector g is incorporated within p to represent subsurface thermal variability, enabling the model to capture non-uniform temperature distributions with depth across different geological conditions.

The number of hidden layers is implicitly defined by the length of each layer-wise neuron configuration, and no additional specification of layer count is required. A nonlinear activation function is applied after each hidden layer to enable the network to capture the complex nonlinear thermal behavior of the coaxial CLGS, while the output layer employs a linear activation function to produce continuous temperature predictions. The final network architecture is selected by jointly considering predictive accuracy, convergence behavior, and computational efficiency. Predictive performance is evaluated using $R^2$, MAE, and RMSE metrics, while training stability is assessed through loss convergence characteristics.

In conventional geothermal heat-transfer PINN models, system behavior is typically described by transient partial differential equations (PDEs) that account for both spatial and temporal variations. However, the governing equations are simplified to ordinary differential equations (ODEs) in this study along the axial coordinate of the wellbore. Because fluid temperature adjusts rapidly during circulation, whereas the surrounding formation temperature evolves gradually due to long-term thermal depletion. As a result, the formation temperature distribution can be approximated as quasi-steady within each operational year, with the dominant variation occurring along the depth direction. Accordingly, the governing equations reduce to one-dimensional axial energy balance equations expressed as ODEs in terms of the spatial coordinate $z$, which are enforced in the PINN framework through the residual loss.

Therefore, the loss function used to train the PINN is constructed as a weighted combination of three components: a data loss term, a governing-equation (ODEs) loss term, and a boundary-condition (BC) loss term. The total loss function is expressed as:

$$\mathcal{L}(\boldsymbol{\theta}) = w_{Data}\mathcal{L}_{Data} + (w_{ODE}\mathcal{L}_{ODE} + w_{BC}\mathcal{L}_{BC}) \qquad (14)$$

where $w_{Data}, w_{ODE}, w_{BC}$ the weighting coefficients control the relative contributions of each term to balance the relative contributions of physical residuals and sparse observations constraints in the loss function, following the reference formulation of physics-informed neural networks. In practice, these weights are tuned to ensure stable convergence and to prevent dominance of any single loss component.

The data loss term, $\mathcal{L}_{\mathrm{Data}}$, is defined to measure the discrepancy between the observable system response and the corresponding model prediction at the wellhead. The governing-equation loss, $\mathcal{L}_{ODE}$, enforces compliance with the heat transfer equations governing the coupled thermal behavior of the inner pipe, annulus, and surrounding formation. The boundary condition loss, $\mathcal{L}_{BC}$, enforces that several physical boundary conditions can be explicitly defined and incorporated into the model loss function.

In practical geothermal operations, temperature measurements are typically available only at the wellhead, while temperature distributions along the wellbore are rarely accessible. As a result, the available observations are limited to the outlet temperature, which serves as the model's primary supervisory signal. Specifically, the predicted inner pipe temperature at the wellhead is enforced to match the observed outlet temperature, expressed as:

$$\hat{T}_{inner}(0) = T_{out} \qquad (15)$$

where $\hat{T}_{inner}(0)$ denotes the temperature predicted by the PINN at the wellhead, and $T_{out}$represents the observed outlet temperature. The corresponding data loss term is defined as

$$L_{data} = \left|\hat{T}_{inner}(0) - T_{out}\right|^{2} \qquad (16)$$

The governing equations used to construct the physics-informed residuals are derived from one-dimensional energy conservation along the wellbore. The depth coordinate $z$ is defined as positive downward. In the coaxial CLGS considered in this study, the working fluid flows downward in the annulus and upward in the inner pipe, leading to opposite signs in the convective heat transport terms of the corresponding energy balance equations.

For the annulus flow, the energy balance accounts for axial convective transport as well as heat exchange with both the surrounding formation and the inner pipe. The governing equation is written as:

$$\mathrm{R_{outer}(z)} = \frac{\mathrm{d}\hat{T}_{\mathrm{outer}}}{\mathrm{dz}} - \frac{\frac{\mathrm{T_{ring}} - \hat{T}_{\mathrm{outer}}}{\mathrm{R_{wall}}} + \frac{\hat{T}_{\mathrm{inner}} - \hat{T}_{\mathrm{outer}}}{\mathrm{R_{pipe}}}}{\dot{\mathrm{m}}\mathrm{c_p}} \qquad (17)$$

| Symbol | Description | Unit |
|---|---|---|
| z | Axial coordinate along the wellbore | m |
| $\mathrm{R_{outer}(z)}$ | Residual of the governing equation for the outer pipe | – |
| $R_{inner}(z)$ | Residual of the governing equation for the inner pipe | – |
| $\hat{T}_{\mathrm{outer}}$ | Predicted temperature of the annulus (outer flow) from the PINN | °C |
| $\hat{T}_{inner}$ | Predicted temperature of the inner tube flow from the PINN | °C |
| $T_{ring}$ | Formation (rock) temperature surrounding the wellbore | °C |
| $R_{wall}$ | Thermal resistance between annulus fluid and surrounding formation ($R_{steel}$ + $R_{cement}$ *or* $R_{steel}$) | K W$^{-1}$ |
| $R_{pipe}$ | Thermal resistance between inner tube and annulus flows | K W$^{-1}$ |

For the inner pipe flow, the energy balance accounts for axial convective transport and heat exchange with the annulus, opposite to the positive $z$-direction. Consequently, the governing equation is written as:

$$R_{\text{inner}}(z) = -\frac{d\hat{T}_{outer}}{\text{dz}} - \frac{\frac{\hat{T}_{inner} - \hat{T}_{outer}}{R_{pipe}}}{\dot{m}c_p} \tag{18}$$

The corresponding ODE residual loss is defined over a set of collocation points $z_f$ sampled along the wellbore:

$$L_{ODE} = \frac{\mathbf{1}}{\boldsymbol{N_f}} \sum_{z_f} (R_{outer}^2 + R_{inner}^2) \tag{19}$$

| Symbol | Description | Unit |
|---|---|---|
| $N_f$ | Number of collocation points used in the PINN training | – |
| $z_f$ | Collocation point used for evaluating the physics residual | m |

To enforce known thermal constraints at the boundaries of the coaxial closed-loop geothermal system. These constraints include prescribed temperatures at the wellhead and physical consistency conditions at the bottom of the well. At the wellhead ( $z = 0$ ), the annulus (inlet) pipe temperature is constrained using the available simulation data. The annulus temperature at the wellhead corresponds to the inlet fluid temperature, the wellhead boundary loss is defined as:

$$\hat{T}_{outer}(0) = T_{in} \tag{20}$$

At the bottom of the well ($z = Z$), a physical continuity condition is imposed to ensure thermal equilibrium between the annulus and the inner pipe. Since the fluid transitions from the downward annulus flow to the upward inner-pipe flow at the bottom, their temperatures are expected to be equal. This condition is enforced through

$$\hat{T}_{inner}(\boldsymbol{Z}) = \hat{T}_{outer}(\boldsymbol{Z}) \tag{21}$$

The total boundary condition loss is obtained by combining the above components as

$$L_{BC} = |\hat{T}_{outer}(0) - T_{in}|^2 + |\hat{T}_{inner}(H) - \hat{T}_{outer}(H)|^2 \tag{22}$$

By jointly considering the data loss and the physics-based loss components, the PINN is guided to satisfy the governing equations and boundary constraints while remaining consistent with the limited wellhead temperature observations. This composite loss formulation allows the network to infer the internal temperature evolution along the wellbore without being explicitly provided with intermediate temperature information. As a result, the trained PINN achieves physically consistent predictions that respect both the observable boundary data and the underlying heat transfer mechanisms of the coaxial closed-loop geothermal system.

## 2.4 ORC Model

To convert the thermal output predicted by the PINN into electrical power generation, an Organic Rankine Cycle (ORC) performance model is introduced and coupled with the geothermal system model. The ORC model evaluates the power generation potential of the predicted geothermal production temperature under realistic ambient conditions, thereby enabling an integrated assessment of geothermal heat extraction and electricity production. The production fluid temperature form wellhead is used as the hot-side inlet temperature to the ORC evaporator. The geothermal reinjection temperature is treated as a prescribed control setpoint that determines the outlet temperature of the geothermal fluid from the ORC system after heat extraction. Together with the specified geothermal mass flow rate, these variables define the available geothermal heat input to the ORC model.

The ORC model employed in this study follows a conventional steady-state thermodynamic cycle framework commonly used in geothermal power generation. The model evaluates the working fluid states at key cycle points, including the condenser outlet, pump outlet, turbine inlet, and turbine outlet, based on prescribed boundary conditions and cycle assumptions [31]. Such approaches have been widely applied in the analysis and optimization of geothermal ORC systems, including parametric studies of operating conditions and working fluids, and have demonstrated good agreement with detailed simulation tools for performance prediction under fixed boundary conditions [32]. Accordingly, the present model adopts this established framework to compute instantaneous gross and net power outputs, as well as thermodynamic efficiency, based on the specified operating conditions. Condenser boundary conditions are defined using historical meteorological data to account for realistic ambient influences on ORC performance. The ambient dry-bulb temperature is used to determine the condensing temperature for air-cooled operation, while the wet-bulb temperature is used for water-cooled operation. These parameters govern the achievable condenser saturation temperature and, consequently, the thermodynamic performance of the ORC system. Hourly weather data are incorporated to capture temporal variability in ambient conditions and their impact on power generation.

The key assumptions and design parameters adopted in the ORC performance model, including evaporator pinch-point temperature differences, condenser approach temperatures and component efficiencies are summarized in Table 2. The well depth and inner wellhead diameter are also included, as they determine the hydraulic length scale and flow area used to estimate the geothermal pumping power required for fluid circulation through the wellbore.

The ORC working fluid is selected based on the range of geothermal production temperatures. A representative refrigerant is selected based on reported industrial practices and performance characteristics, with reference to publicly available Ormat system reports [33] and established ORC design guidelines presented in the World Geothermal Congress [34].

Thermophysical properties of the working fluid are evaluated using the CoolProp library [35], which provides the necessary thermodynamic properties, including saturation pressure, enthalpy, entropy, density, and vapor quality at relevant cycle states. Historical meteorological data required for the ORC simulation are obtained using the Meteostat package [36]. Hourly records of ambient dry-bulb and wet-bulb temperatures are used to represent site-specific atmospheric conditions and their temporal variability.

Table 2. Key assumptions and parameters of the ORC performance model

| Parameter | Symbol / Value | Description |
|---|---|---|
| | **Geothermal System** | |
| Geothermal inlet temperature | $T_{geo,in}$ | Provided by PINN prediction (wellhead production temperature) |
| Geothermal reinjection temperature | $T_{geo,out}$ | Prescribed control setpoint with process variability |
| Geothermal mass flow rate | $\dot{m}_{geo}$ | Geothermal fluid mass flow rate |
| Well depth | $L_{well}$ | Depth used for geothermal pumping pressure calculation |
| Well inner diameter | $D_{inner}$ | Inner diameter used to calculate flow area and pressure loss |
| | **ORC design** | |
| Working fluid | *Isopentane/ Isobutane* | Selected based on geothermal conditions |
| Cooling type | Air / Water cooled | Condenser cooling method |
| Evaporator pinch-point temperature difference | $\Delta T_{pinch} = 15°C$ | Minimum temperature difference in evaporator |
| Condenser approach temperature (air cooling) | $\Delta T_{app} = 15°C$ | Temperature difference above ambient air |
| Subcooling | $\Delta T_{sub} = 5°C$ | Subcooling at condenser outlet |
| Superheat | $\Delta T_{sup} = 3°C$ | Superheat at turbine inlet |
| Critical temperature margin | $\Delta T_{crit} = 5°C$ | Ensures subcritical operation |
| | **Performance assumptions** | |
| Isentropic efficiency | $\eta_{turb} = 0.80$ | Nominal value |
| Pump isentropic efficiency | $\eta_{pump} = 0.75$ | Assumed constant |
| Mechanical efficiency | $\eta_{mech} = 0.85$ | Shaft efficiency |
| Generator efficiency | $\eta_{gen} = 0.90$ | Electrical conversion efficiency |

Only key design and operating parameters specific to the present study are listed, while standard thermophysical properties and auxiliary assumptions follow typical ORC design practices reported in the literature. In particular, the evaporator pinch-point temperature difference is commonly selected in the range of 10–20°C, with 15°C representing a typical optimal value. Similarly, condenser approach temperature, subcooling, and superheating values are chosen within standard engineering ranges to balance system efficiency and heat exchanger size [37][38].

After completing the thermodynamic calculations, the cycle energy balance is evaluated to determine system performance. The turbine power output, pump consumption, and net electrical power are obtained from the enthalpy differences between the thermodynamic state points. The resulting quantities, including net power output, thermal efficiency, and state properties, are returned as the outputs of the ORC simulation and subsequently used as inputs to the economic evaluation model for assessing the overall system performance.

## 2.5 Economic Analysis Model

To evaluate the feasibility of redeveloping abandoned oil wells as closed-loop geothermal power systems, an economic model is developed by coupling the ORC-derived power output with key economic parameters, including capital expenditure (CAPEX) and operating expenditure (OPEX). The analysis is conducted on a single-well basis, where both power generation and associated costs are estimated for an individual well converted into a closed-loop geothermal system. Considering long-term operation, key economic indicators including levelized cost of energy (LCOE), net present value (NPV), and discounted payback period (DPP) [21], [22] are calculated to assess the financial viability of the system. The LCOE represents the ratio of the net present value of total system costs to the net present value of the total electricity generated, the NPV is used to evaluate the overall profitability of the geothermal power project. It represents the difference between the discounted revenues and the discounted costs over the lifetime of the project and the DPP time required for the project to recover its initial investment when the time value of money is considered.

The cost parameters and electricity price used in the model are obtained from published literature and publicly available data sources related to geothermal well retrofitting, geothermal power systems, and ORC-based power generation. The capital expenditure (CAPEX) primarily includes the surface ORC system, well redevelopment or drilling costs, and associated infrastructure required for converting abandoned oil wells into closed-loop geothermal systems. The cost structure is referenced from the techno-economic assessment of the Eavor-Loop system [24], which provides representative estimates of development and installation costs. In addition to the initial investment, annual operating expenditure (OPEX) is considered, including maintenance of the ORC system, monitoring, and routine operational activities. Thermodynamic and economic characteristics of the ORC system are adopted from previous thermo-economic studies[21].

Based on these inputs, the economic indicators are calculated by the following formulations. Levelized Cost of Energy:

$$LCOE = \frac{CAPEX + \sum_{t=1}^{N} \frac{OPEX_t}{(1+r)^t}}{\sum_{t=1}^{N} \frac{E_t}{(1+r)^t}} \tag{23}$$

| Symbol | Description | Unit |
|---|---|---|
| $CAPEX$ | Total capital expenditure for system installation | USD |
| $OPEX_t$ | Operating and maintenance cost in year $t$ | USD yr$^{-1}$ |
| $E_t$ | Electricity generated in year $t$ | kWh yr$^{-1}$ |
| t | Year index | - |
| N | Project lifetime | years |
| r | Discount rate | - |

Net Present Value:

$$NPV = -CAPEX + \sum_{t=1}^{N} \frac{R_t - OPEX_t}{(1+r)^t} + \frac{SV - DC}{(1+r)^N} \tag{24}$$

$$R_t = P_t E_t$$

| Symbol | Description | Unit |
|---|---|---|
| $R_t$ | Revenue generated in year $t$ | USD $yr^{-1}$ |
| $P_t$ | Electricity price in year $t$ | USD |
| $SV$ | Salvage value | USD |
| $DC$ | Decommissioning cost | USD |

A positive NPV indicates that the project is economically profitable under the given assumptions.

Discounted Payback Period:

$$DPP = \min\left\{t \in [1,N] \,\middle|\, \sum_{i=1}^{t} \frac{R_i - OPEX_i}{(1+r)^i} \geq CAPEX\right\} \quad (25)$$

where $t$ denotes the year index over which discounted cash flows are accumulated, and $N$represents the total project lifetime. The operator $\min(\cdot)$ identifies the earliest year $t$ at which the cumulative discounted cash flow equals or exceeds the initial investment.

Unlike LCOE and NPV, which assess economic performance over the entire project lifetime, the DPP focuses on the time required to recover the initial investment.

By integrating the PINN-predicted geothermal production temperature, the ORC-based power generation model, the economic evaluation is performed through a year-by-year simulation over a project lifetime of $N$ years. Within each year, calculations are conducted on an hourly basis (8,760 hours). At each time step, the geothermal outlet temperature is updated based on the operating time and used to determine the corresponding net power output. The hourly results are accumulated to obtain the annual electricity generation.

At the end of each year, the annual electricity generation and operating costs are used to compute the net cash flow. The discounted cash flows are then evaluated to calculate the economic indicators. The NPV and LCOE are determined over the full project lifetime $N$, while the DPP is obtained by sequentially accumulating discounted cash flows until the cumulative value becomes non-negative.

If the cumulative discounted cash flow continues to decrease, indicating persistent negative returns, the calculation is terminated early, and the project is considered to have no feasible payback period within the evaluation horizon.

## 3. IMPLEMENTATION AND RESULTS

This section presents the implementation and results of the proposed framework for closed-loop geothermal systems. First, the simulation dataset is introduced to illustrate the operating conditions and thermal behavior of the system. Next, the performance of the physics-informed neural network (PINN) is evaluated, including model validation and hyperparameter tuning to identify suitable model configurations. Subsequently, the predicted thermal outputs are integrated with the Organic Rankine Cycle (ORC) model to estimate power generation performance. Finally, a techno-economic analysis is conducted to evaluate system feasibility based on key economic indicators, including the discounted payback period.

### 3.1 Synthetic Data Simulation

based on historical geothermal data from Iceland, ensuring that the simulated subsurface conditions are physically representative of realistic geothermal environments. The ranges of randomly sampled input parameters are defined according to reported field data and are summarized in Table 3. Simulation parameters including formation properties, wellbore configuration, and working fluid conditions. These parameters include operational conditions, well geometry, and rock thermal properties.

Table 3. Simulation parameters including formation properties, wellbore configuration, and working fluid conditions

| Parameter | Value | Unit |
|---|---|---|
| Rock thermal conductivity | 1.5 – 2.5 | $W\ m^{-1}\ K^{-1}$ |
| Rock thermal diffusivity | $5\times10^{-7}$ – $8\times10^{-7}$ | $m^2\ s^{-1}$ |
| Geothermal gradient (baseline) | 50 – 150 | $°C\ km^{-1}$ |
| Gradient Variation | ±15% | – |
| Inlet temperature | 60 – 80 | °C |
| Inner tube diameter | 0.4 – 1.0 | m |
| Mass flow rate | 2.0 – 8.0 | $kg\ s^{-1}$ |
| Operation time | 1-30 | year |

| Parameter | Value | Unit |
|---|---|---|
| Well depth | 2000 – 8000 | m |

Specifically, the inlet temperature, pipe diameter, mass flow rate, operation time, and well depth is randomly sampled within predefined ranges. In addition, the thermal conductivity and diffusivity of the surrounding formation are sampled to reflect variability in rock properties. The geothermal gradient is modeled as a segmented profile along the depth, where each segment is perturbed around a base gradient using a controlled noise level to represent subsurface heterogeneity.

To illustrate the generated dataset, a representative sample is selected with the following parameters: inlet temperature of 65.7 °C, pipe diameter of 0.819 m, mass flow rate of $6.03\ \mathrm{kg\ s^{-1}}$, well depth of 5158 m, and operation time of 30 years. The rock thermal conductivity and diffusivity are $2.0\ \mathrm{W\ m^{-1}K^{-1}}$ and $6.50 \times 10^{-7}\mathrm{m^2 s^{-1}}$, respectively. The base geothermal gradient is $87.5°\mathrm{C\ km^{-1}}$, with segment-wise variations ranging from approximately 79 to 95 $°\mathrm{C\ km^{-1}}$.

Based on these geothermal conditions, the simulated wellbore temperature profile after 30 years of operation is shown in Figure 3 The annulus temperature increases with depth as the fluid absorbs heat from the surrounding formation, while the inner pipe temperature represents the upward transport of thermal energy toward the surface. The temperature difference between the two flow paths is governed by local heat transfer intensity and flow conditions.

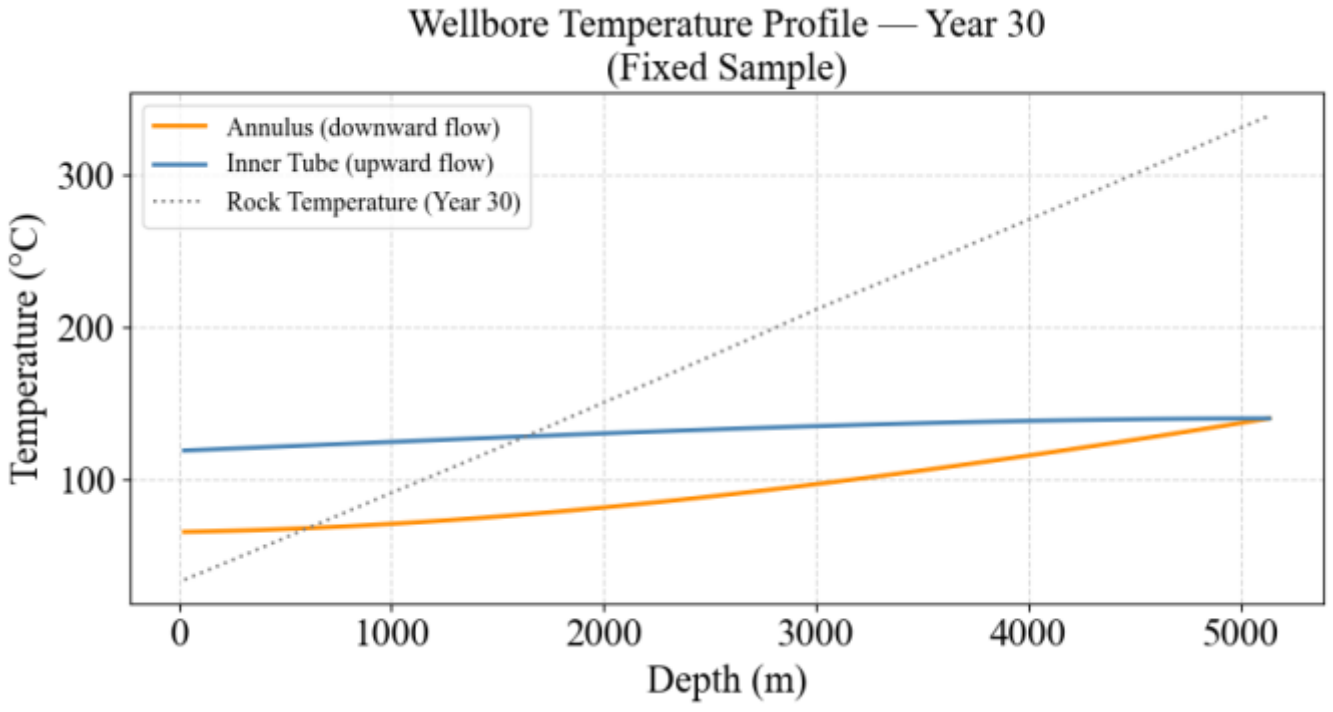

Figure 3. Wellbore temperature profile

In addition to the wellbore temperature evolution, the thermal behavior of the surrounding formation is further illustrated in Figure 4, which shows the temporal and depth dependent cooling of the rock. The results indicate that thermal depletion is most pronounced in the vicinity of the wellbore, while deeper regions retain higher temperatures over time. This demonstrates that heat extraction is initially dominated by the near-well region, followed by gradual heat replenishment from the surrounding formation through thermal conduction.

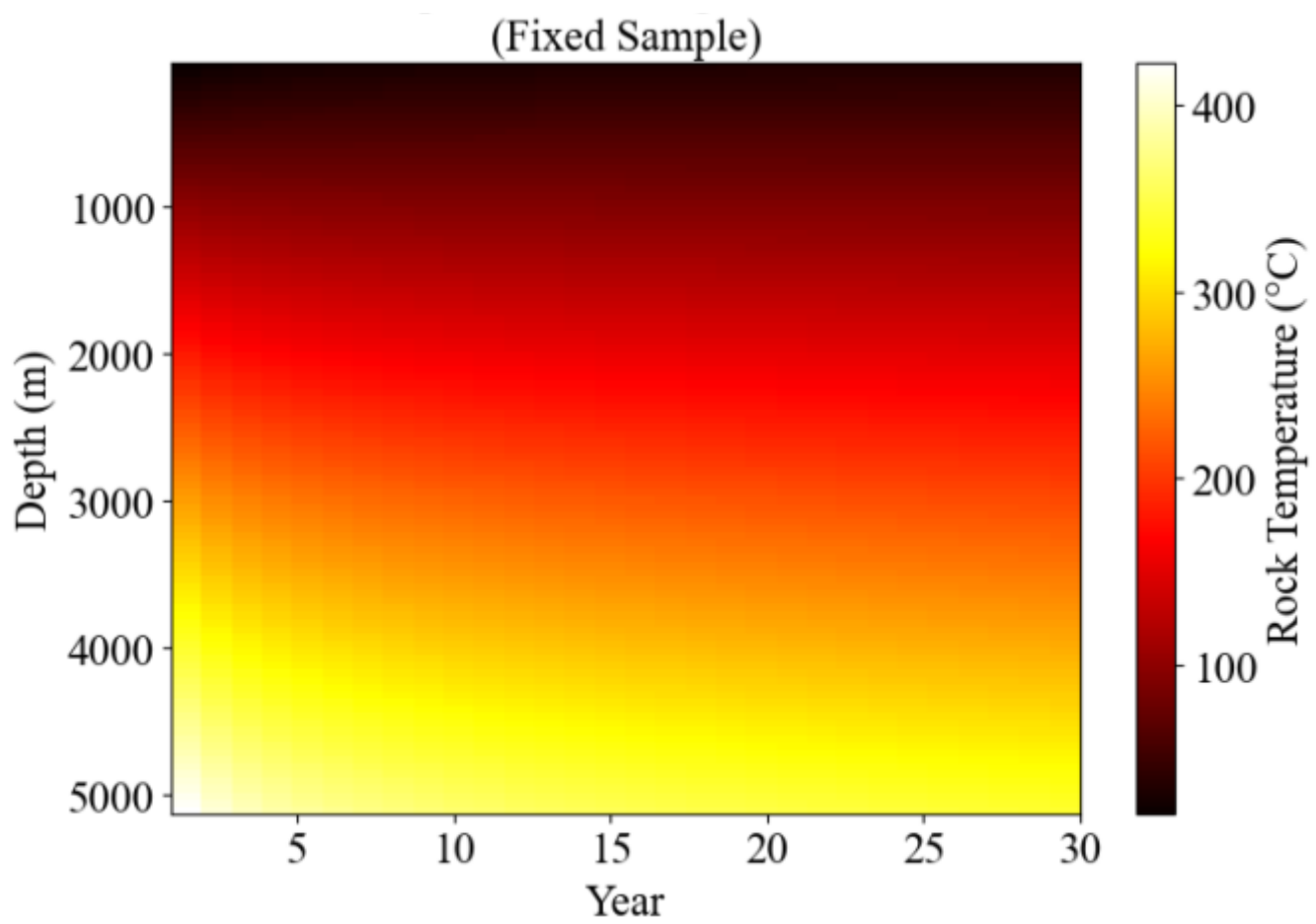

**Figure 4**. Heatmap of rock temperature evolution around the well

The long-term impact of thermal depletion is further reflected in the evolution of the outlet temperature, as shown in **Figure 5**. The outlet temperature exhibits a rapid decline during the initial years of operation, followed by a gradual stabilization over time. This behavior is consistent with the reduction in temperature difference between the circulating fluid and the surrounding formation as heat is extracted.

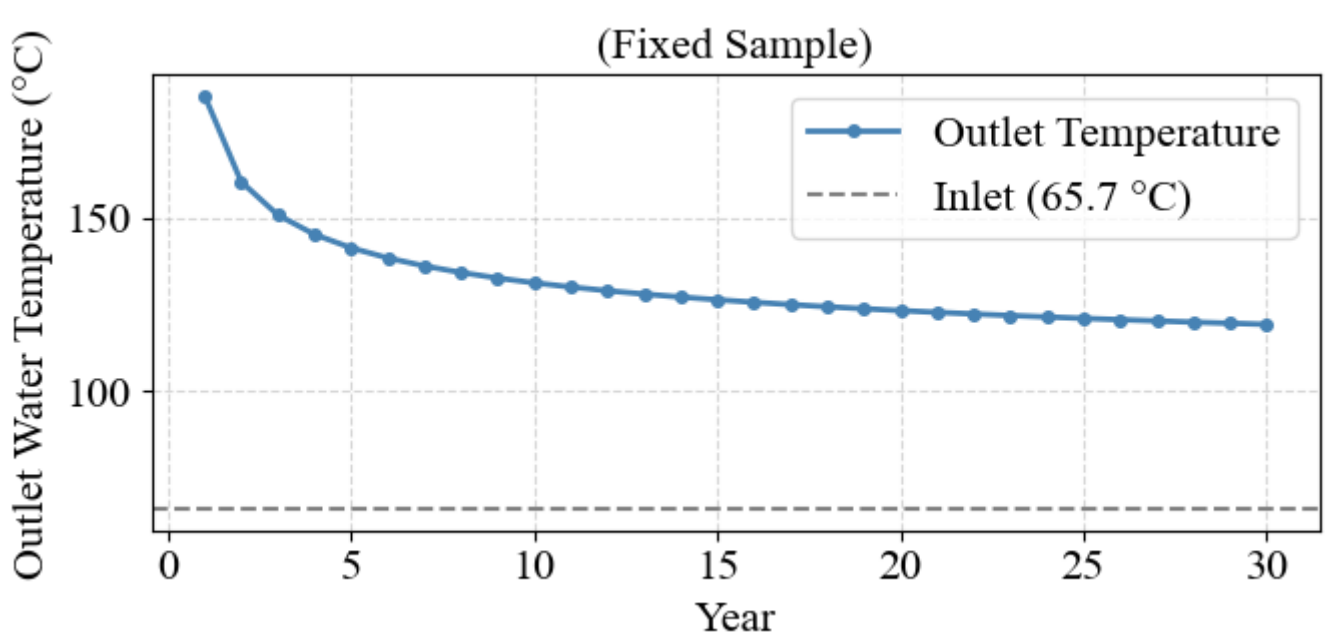

**Figure 5**. Annual outlet temperature over 30 years

Although the full wellbore temperature distributions and long-term temporal evolution are available from the simulation framework, only limited observable data are assumed to be accessible for PINN training. Thus, the model is provided with known wellbore parameters along with measurable wellhead temperatures, including the inner pipe and annulus temperatures, and the corresponding operating time. This setup reflects a practical scenario in which downhole temperature measurements are not readily available, while surface measurements can be continuously monitored. The use of such limited observations is therefore intended to emulate realistic field conditions, where only partial system information can be obtained.

It is assumed that, with future deployment of monitoring systems, sufficient operational data can be collected to support model training. Under this assumption, the PINN leverages both sparse observational data and governing physical constraints to reconstruct the full temperature distribution along the wellbore.

### 3.2 PINN Model Training and Performance

To investigate the dependence of model performance on data availability, the PINN is trained using datasets of varying sizes, including 100, 500, 1000, and 4000 samples. This setup enables a systematic evaluation of the model's ability to learn from sparse observations and to assess its generalization capability under limited data conditions.

For hyperparameter optimization, the Optuna [39] is employed to identify an optimal network configuration. The search space includes the number of hidden layers, hidden layer width, activation function, and learning rates for both the model parameters and the adaptive loss balancing scheme. A uniform architecture is adopted for each trial, where all hidden layers share the same number of neurons. The ranges of the hyperparameters considered in this study are summarized in Table 4.

The optimization is conducted using a Tree-structured Parzen Estimator (TPE) [40] sampler with a total of 30 trials, and a median pruning strategy is applied to terminate unpromising trials during training. Model performance is evaluated using a weighted mean absolute error (MAE), which combines the wellhead prediction error and the full-depth temperature profile error. The profile error is computed by comparing PINN predictions with pre-computed simulation results, ensuring that both boundary accuracy and internal thermal consistency are considered in the model selection process.

Table 4. Key hyperparameters range for PINN optimization test

| Category | Parameter | Range / Values |
|---|---|---|
| Architecture | Number of hidden layers | 2 – 6 |
| | Neurons per layer | 64, 128, 256 |
| | Activation function | tanh, GELU, SiLU |
| Optimization | Learning rate (model) | $(10^{-4}) \sim (5 \times 10^{-4})$ |
| | Learning rate (loss balancer) | $(5 \times 10^{-4}) \sim (5 \times 10^{-3})$ |
| Evaluation | Objective metric | Weighted MAE (wellhead + profile) |

The prediction results indicate that the PINN model is capable of capturing the overall temperature trends along the wellbore even with a limited training dataset. As shown in Figure 6, the model trained with 100 samples is able to predict the general shape of the temperature profile, although noticeable deviations remain, particularly in regions with stronger thermal gradients.

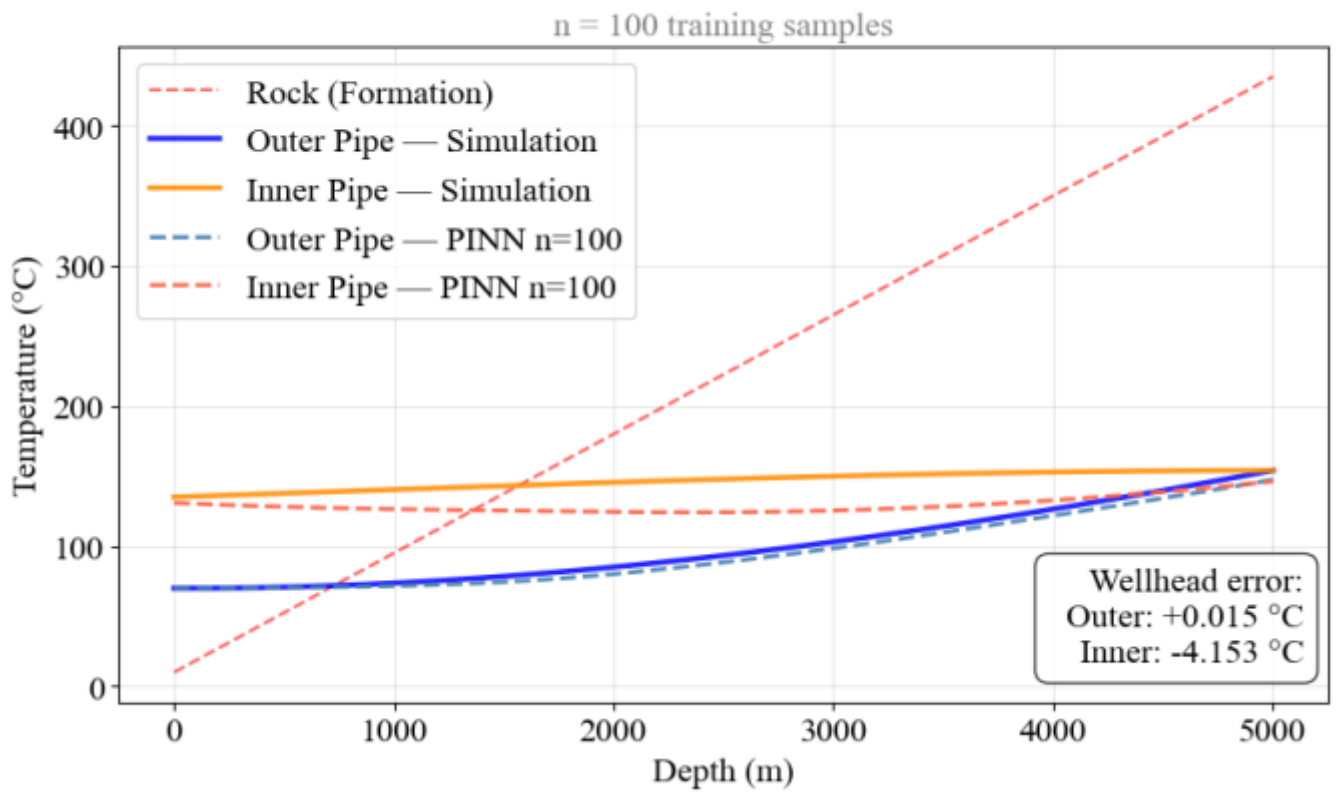

**Figure 6**. Temperature profile - PINN (n = 100) prediction vs simulation

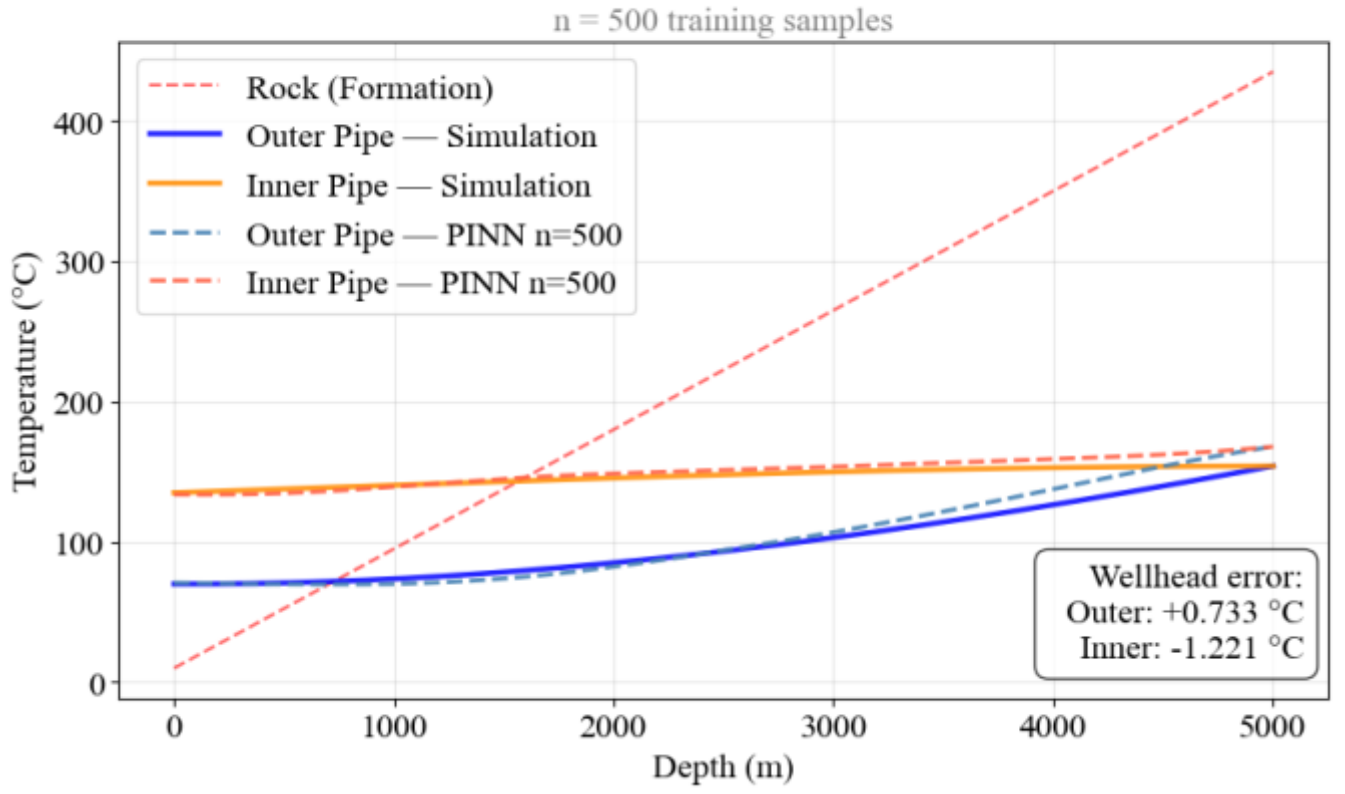

**Figure 7**. Temperature profile - PINN (n = 500) prediction vs simulation

As shown in Figure 7, when the training dataset is increased to 500 samples, a substantial improvement in prediction accuracy is observed. Both the annulus and inner pipe temperature profiles are more closely aligned with the simulation results, and the overall error is significantly reduced.

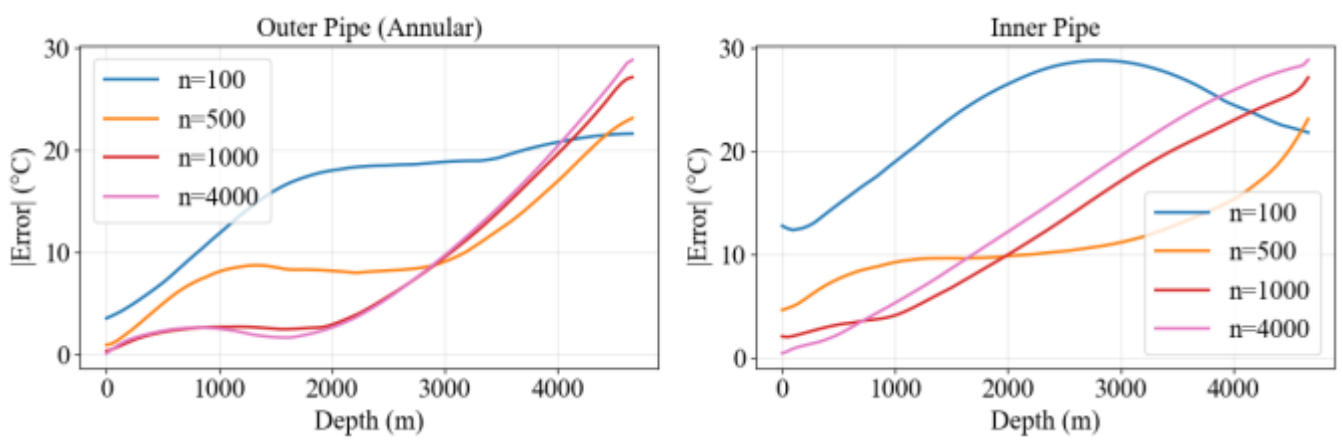

**Figure 8**. Mean error in annulus and inner pipes - PINN prediction vs simulation

However, further increasing the dataset size to 1000 and 4000 samples yields only marginal improvements. As illustrated in Figure 8, the reductions in MAE become relatively small beyond 500 samples, indicating diminishing returns in model performance with additional data. The error distribution is further evaluated using 50 randomly selected test samples, with the results presented as boxplots in Figure 9. Consistent with previous observations, both the median error and the spread of the distribution decrease as the dataset size increases, indicating improved prediction stability and robustness. These results further confirm that increasing the dataset size beyond 500 samples provides limited additional improvement in model performance.

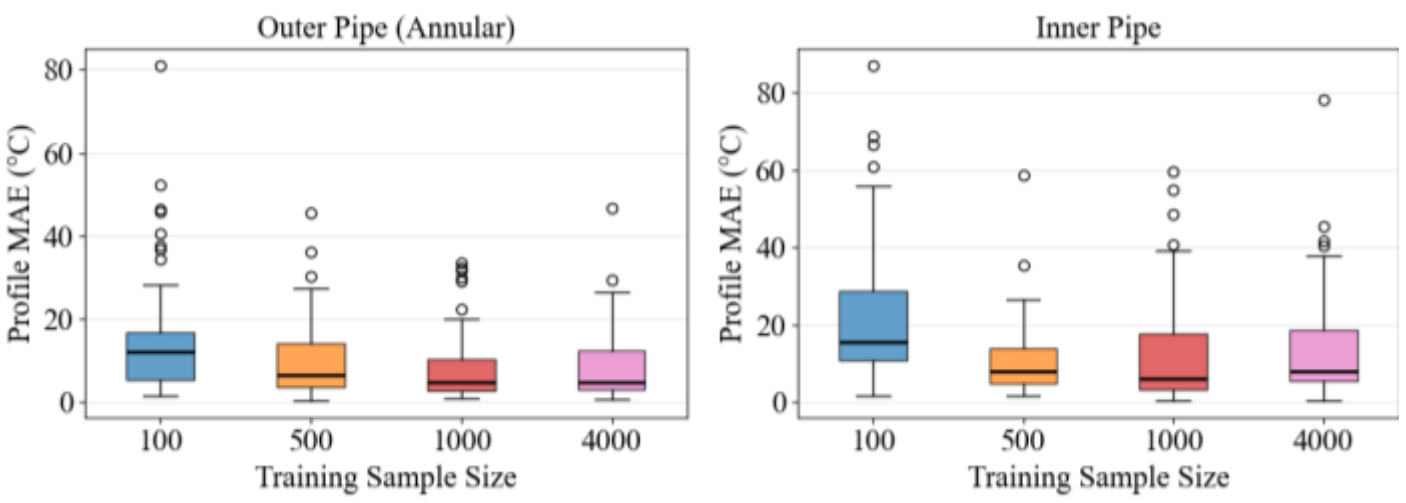

**Figure 9**. CLGS temperature profile MAE distribution 50 test samples

The above results provide a comprehensive evaluation of the PINN model in reconstructing the full wellbore temperature distribution. However, from a practical application perspective, only limited measurements are typically available. In particular, the ORC model relies solely on the predicted inner pipe outlet temperature at the wellhead as its input. Therefore, while the full-depth prediction performance is important for understanding the model behavior, it is also necessary to assess the prediction accuracy at the wellhead, which directly impacts the downstream ORC performance. The corresponding results are presented as Figure 10 and Figure **11** indicate that increasing the number of training samples significantly improves the prediction accuracy at the wellhead. In particular, both MAE and RMSE decrease consistently as the dataset size increases, demonstrating that the model is effectively constrained by additional training data in terms of wellhead temperature prediction.

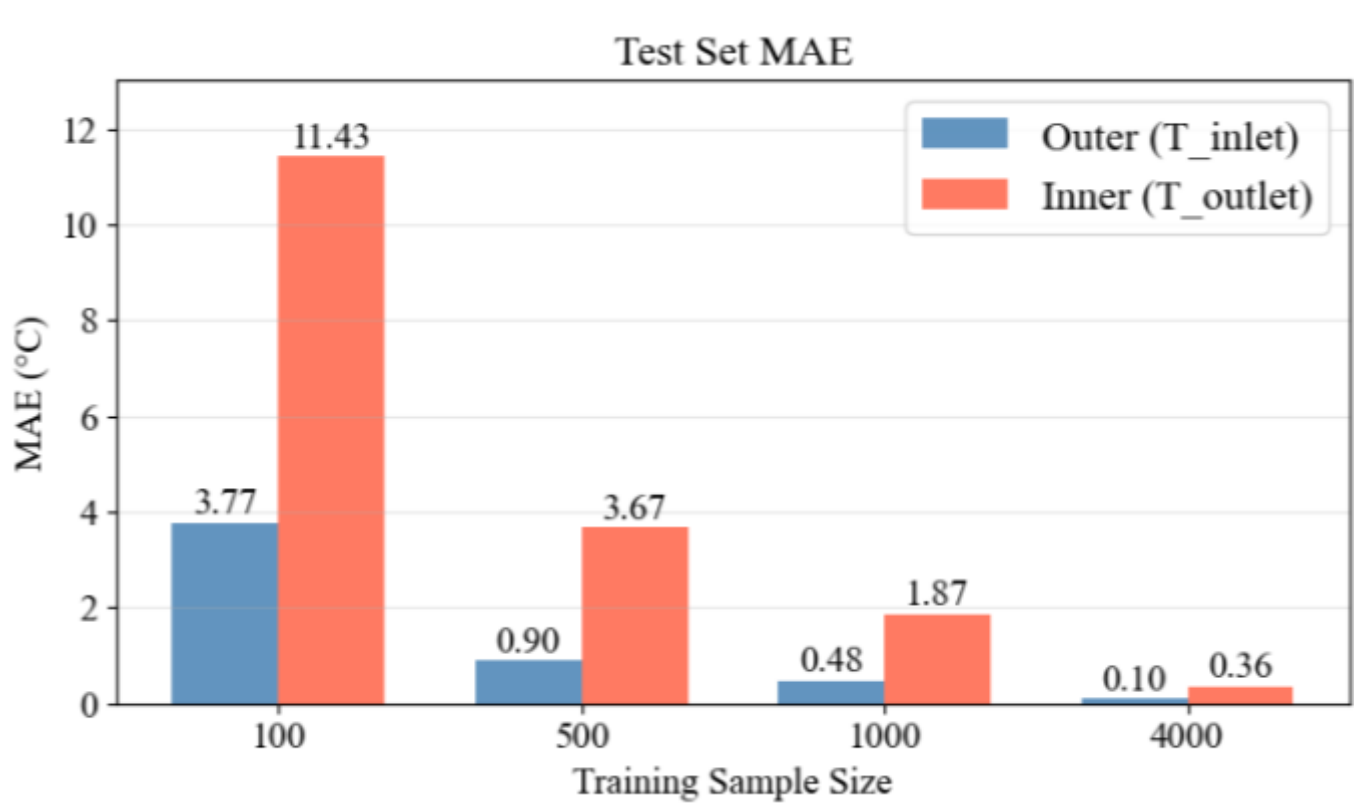

**Figure 10**. Test set MAE of wellhead temperature predictions

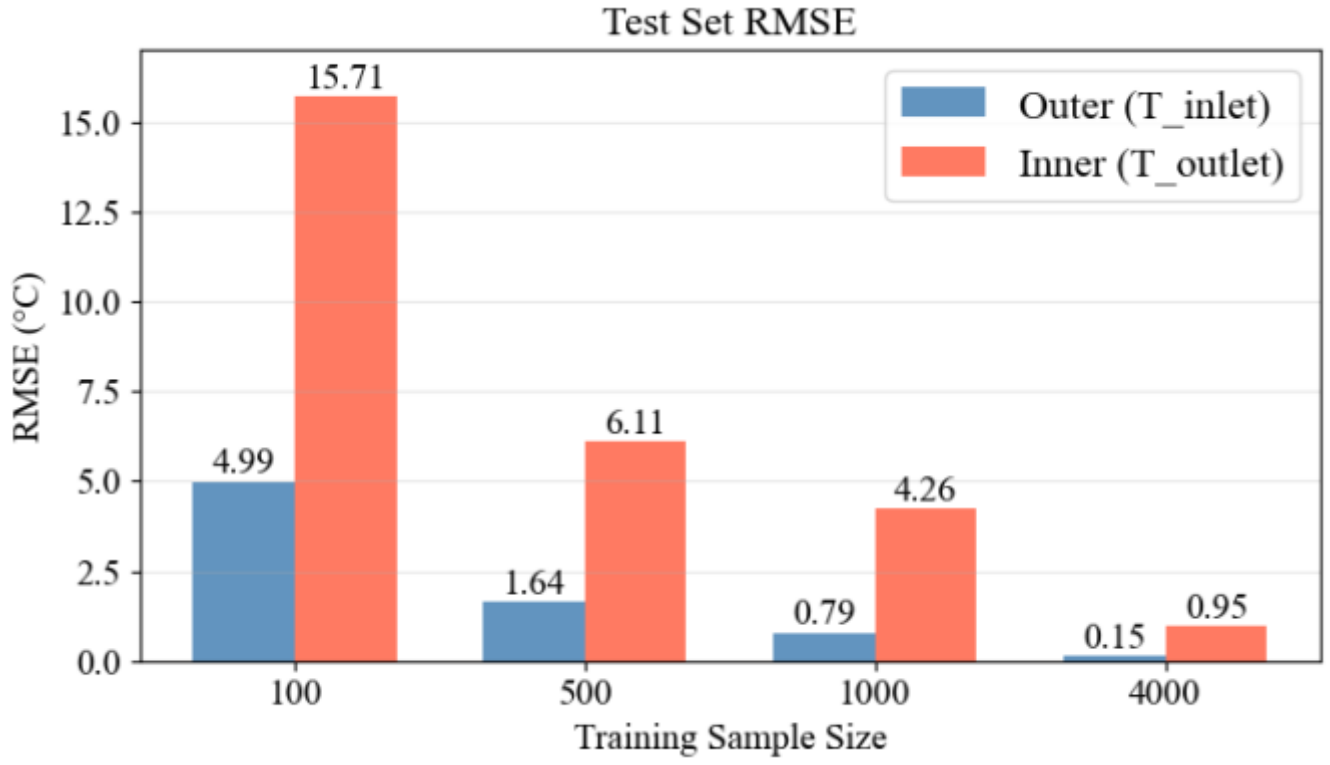

**Figure 11**. Test set RMSE of wellhead temperature Predictions

Overall, a PINN model trained by dataset size of 500 samples is sufficient for accurate full wellbore temperature reconstruction, while the highest accuracy in wellhead temperature prediction is using 4000 samples. Given that the ORC model relies on the wellhead outlet temperature, the model outputs with 4000 samples is adopted for subsequent ORC calculation.

### 3.3 ORC Performance Analysis

Following the wellbore temperature prediction, the ORC performance is evaluated using the wellhead temperature obtained from the selected PINN model along with the corresponding input parameters listed in Table 5, based on typical operating conditions and parameter ranges representative of Icelandic geothermal properties [27] [28].

The ORC system settings and operational variations are summarized in Table 6. The selected working fluid and cooling configuration are defined based on industrial practices reported by Ormat [41] and the ambient input for the ORC model is derived from hourly air temperature data for the year 2020, obtained from the nearest available weather station to Reykjavik, Iceland. While the operational variations are specified according to reported ORC experimental data [20]. These variations are evaluated using the Root-Sum-Square (RSS) method in accordance with the ASME PTC 19.1 standard [42], with an acceptable uncertainty level limited to 5%, and the first-year performance results indicate that the annual average net power output is $W_{\text{net}} = 269.60 \pm 12.54\text{kW}$ (4.65 %), while the net thermal efficiency is $\eta_{\text{net}} = 12.90\% \pm 0.53\%$ (4.12%).

Table 5. Wellbore and geothermal parameters

| Parameter | Value | Unit |
|---|---|---|
| Rejection Temperature | 70.0 | °C |
| Pipe diameter | 0.5 | m |
| Mass flow rate | 5.0 | $\text{kg s}^{-1}$ |
| Well depth | 5000 | m |
| Operation time | 1 ~ 30 | years |
| Rock thermal conductivity | 1.8 | $\text{W m}^{-1}\text{ K}^{-1}$ |
| Rock thermal diffusivity | $(7.83 \times 10^{-7})$ | $\text{m}^2\text{ s}^{-1}$ |
| Geothermal gradient | 85.0 (average at the first year | $°\text{C km}^{-1}$ |

Table 6. ORC system setting and uncertainty assumptions

| Parameter | Value | Unit | Description |
|---|---|---|---|
| Working fluid | Isopentane | - | Suitable for moderate temperature |
| Cooling type | Air | - | Air-cooled condenser |
| Capacity factor | 0.90 | - | Annual operating utilization |
| Temperature noise | 0.3 | °C | Sensor measurement error |
| Flow rate noise | 1% | – | Mass flow measurement uncertainty |
| Pressure noise | 0.5% | – | Pressure sensor uncertainty |
| Turbine efficiency uncertainty | 0.015 | – | Efficiency fluctuation |
| Pump efficiency uncertainty | 0.015 | – | Efficiency fluctuation |
| Pinch temperature noise | 1.0 | °C | Heat exchanger variability |
| Approach temperature noise | 1.5 | °C | Condenser variability |
| Control temperature noise | 0.5 | °C | Operational control fluctuation |

To verify the corresponding thermodynamic behavior of the ORC calculation, the working fluid cycle is illustrated in **Figure 12** and 13. As observed in **Figure 12**, the cycle follows the expected thermodynamic path reported in previous ORC studies [42], where the four principal states are identified. Starting from the condenser outlet, the working fluid is in a low-pressure liquid state. It is then pressurized by the pump to a higher pressure at the evaporator inlet, with a relatively small change in enthalpy.

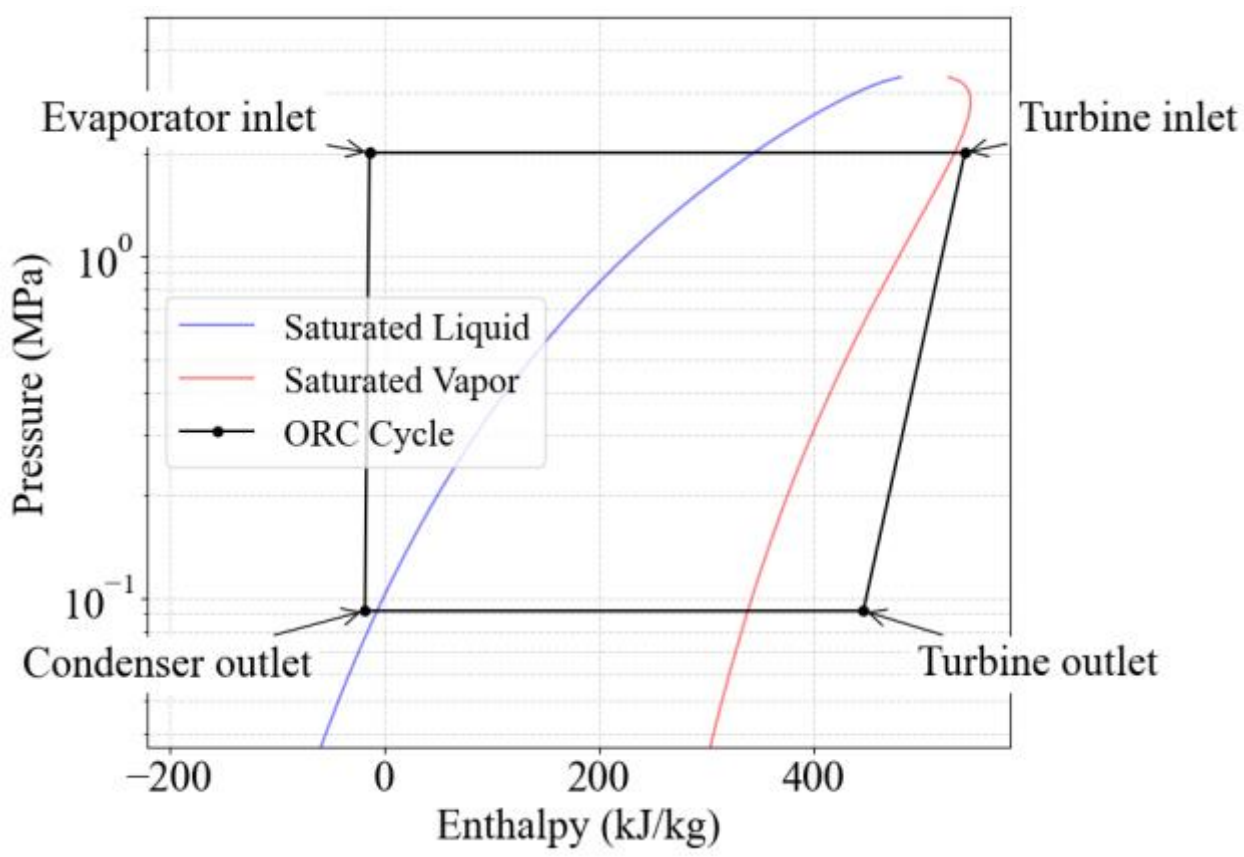

**Figure 12**. P–h Diagram of the ORC Cycle Using Isopentane

Heat is added in the evaporator at approximately constant pressure, increasing the fluid enthalpy from a compressed liquid to a superheated vapor at the turbine inlet. The high-pressure vapor then expands through the turbine, resulting in a pressure drop and a decrease in enthalpy at the turbine outlet. Finally, the fluid is condensed at low pressure, returning to the saturated liquid state and completing the cycle. **Figure 13** further confirms that the cycle remains within the thermodynamically feasible domain, with heat addition occurring at nearly constant temperature and entropy increasing during turbine expansion.

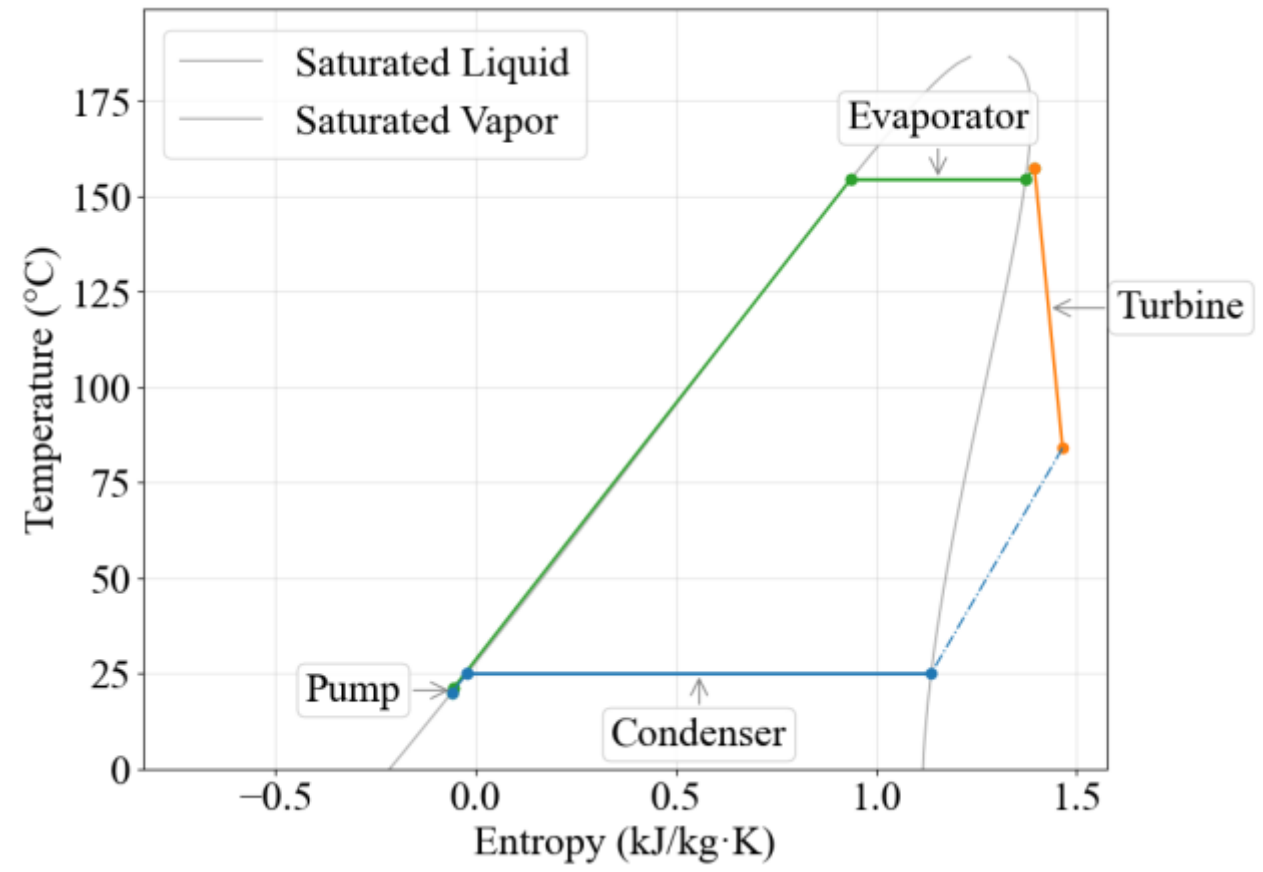

**Figure 13**. T–s Diagram of the ORC Cycle Demonstrating Thermodynamic Consistency (Isopentane)

Having confirmed the thermodynamic consistency of the ORC cycle, the system performance over time is evaluated in terms of net power output and electricity generation. The geothermal outlet temperature is predicted from the PINN model as a function of the given operating year rather than hourly basis. This predicted temperature is then combined with hourly ambient temperature data and used as input to the ORC model to compute the power generation.

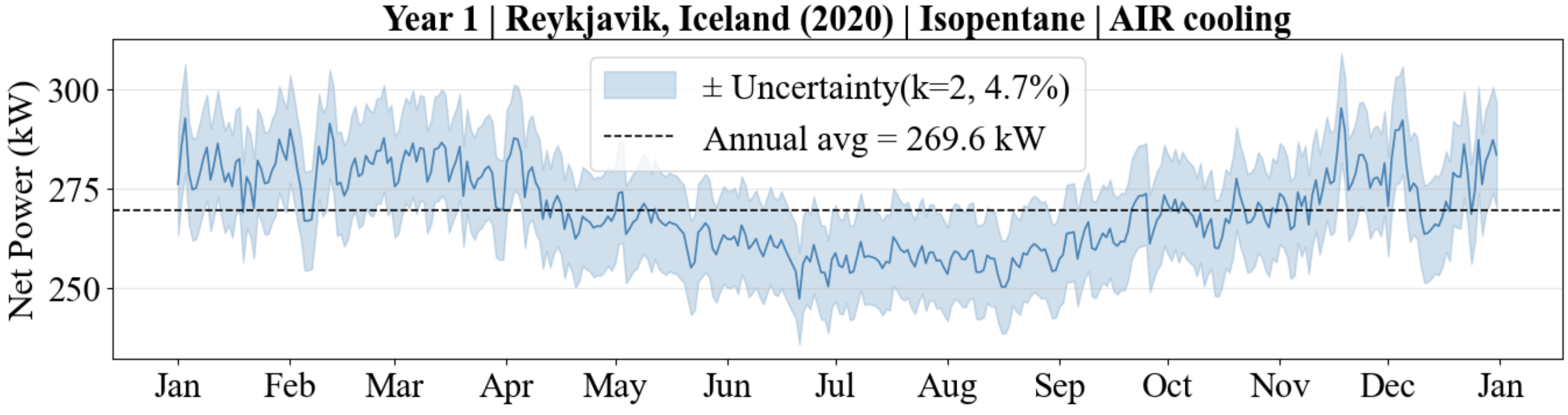

Figure 14. Daily ORC net power output with uncertainty in the first year

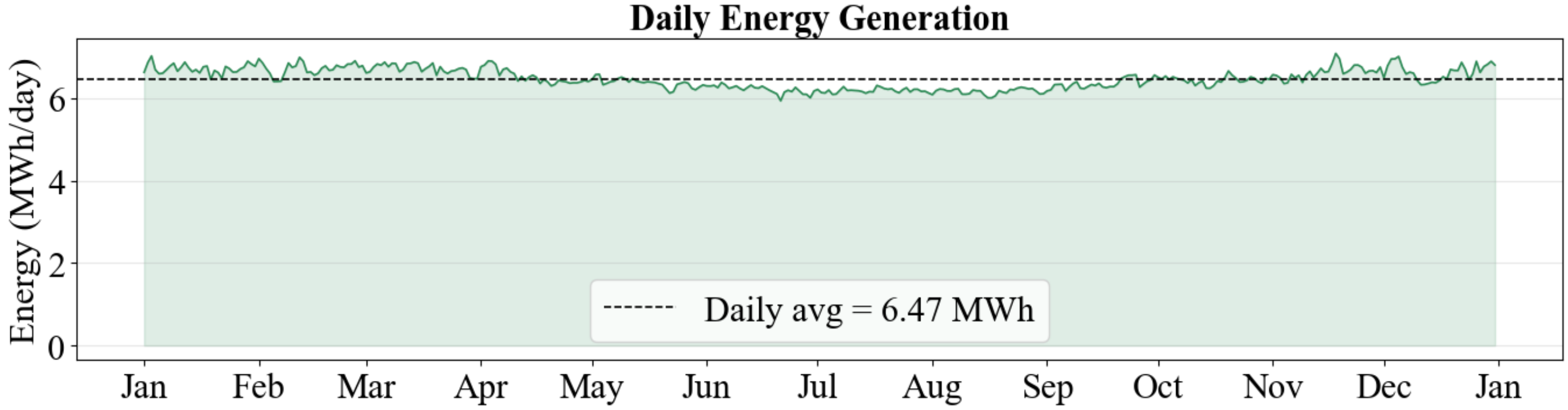

Figure 15. Daily ORC energy generation in the first year

Figure 14 shows the hourly net power output together with the uncertainty band and exhibits moderate seasonal variation, with higher values observed during colder periods and reduced output during warmer months. While Figure 15**Error! Reference source not found.** presents the corresponding daily energy generation and summarizes the daily energy production, indicating a relatively stable generation profile with an average of 6.47 MWh per day. The limited fluctuation in daily output suggests that the system maintains consistent performance despite variations in environmental conditions and operational uncertainties.

Based on the calculated hourly power output, the annual electricity generation is used as the primary input for the subsequent economic evaluation

## Economic Evaluation

The economic performance of the system is evaluated based on the calculated annual electricity generation, using a set of cost parameters categorized into capital expenditure (CAPEX) and operating expenditure (OPEX). The key economic indicators, including the levelized cost of energy (LCOE), net present value (NPV), and discounted payback period (DPP), have been defined in the methodology section and are evaluated here to assess the financial feasibility of the system.

The economic and financial inputs are adopted from published literature and industrial reports and are summarized in Table 7. The parameter values are selected within the reported ranges from the literature and can be adjusted as input variables for different economic scenarios.

Table 7. Economic Parameters Adopted in This Study

| Parameter | Value Range | Unit | Reference |
|---|---|---|---|
| Capital Expenditure (CAPEX) | | | |
| Workover cost | 350,000 – 400,000 | USD well$^{-1}$ | Beckers et al. (2022) [24] |
| Well conversion cost | 700,000 – 900,000 | USD well$^{-1}$ | Beckers et al. (2022) (pipe-in-pipe CLGS) [24] |
| Surface plant cost | 600,000 – 750,000 | USD | DOE GeoVision / IRENA [43], [44] |
| Pipeline cost | 50,000 – 80,000 | USD | Beckers et al. (2022) [24] |
| Contingency factor | 18% – 20% | – | DOE GeoVision [43] |
| Operating Expenditure (OPEX) | | | |
| Fixed O&M cost | 20,000 – 35,000 | USD yr$^{-1}$ | DOE GeoVision [43] |
| Variable O&M cost | 0.001 – 0.002 | USD kWh$^{-1}$ | DOE GeoVision / IRENA [43], [44] |
| End-of-Life Parameters | | | |
| Salvage value | 30,000 – 50,000 | USD | Beckers et al. (2022) [24] |
| Decommissioning cost | 150,000 – 180,000 | USD | DOE GeoVision [43] |
| Financial Parameter | | | |
| Annual price escalation | 1.5% – 2.0% | Per year | EIA long-run forecast [45] |
| Discount rate | 6–12 % | Per year | |
| Init electricity rice | 0.05–0.12 | USD kWh$^{-1}$ | EIA long-run forecast [45] |

Under the base-case economic assumptions, the system does not achieve financial viability over the project lifetime. The LCOE remains consistently higher than the projected electricity price, indicating that the cost of electricity generation exceeds the potential revenue throughout the operation period. This trend is primarily attributed to the gradual decline in thermal performance which leads to reduced annual electricity generation over time, as shown in Figure 16.

From an investment perspective, the NPV remains negative throughout the entire project lifetime, and the DPP is not achieved within 25 years. Although the system generates positive cash flow each year, these returns are not sufficient to recover the initial capital investment.

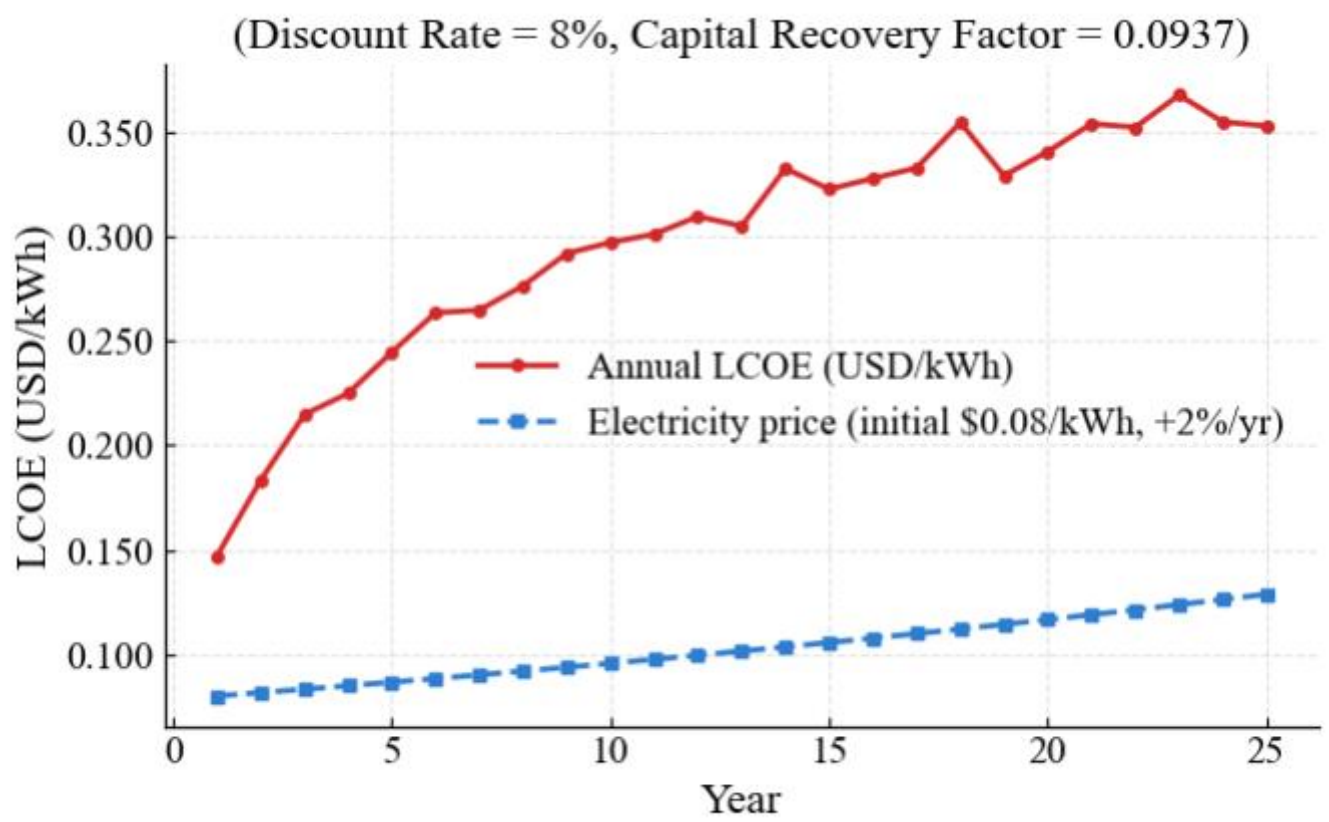

Figure 16. Annual Levelized Cost of Electricity (LCOE)

The trend of the discounted cash flow provides further insight into this behavior. The annual discounted cash flow gradually decreases over time due to both the decline in system performance and the effect of discounting. Meanwhile, the cumulative discounted cash flow continues to increase, but at a slower rate, and never reaches the level of the initial capital cost. Overall, these results in Figure 17 indicate that the economic performance of the system is limited by high upfront investment and insufficient electricity generation over the long term.

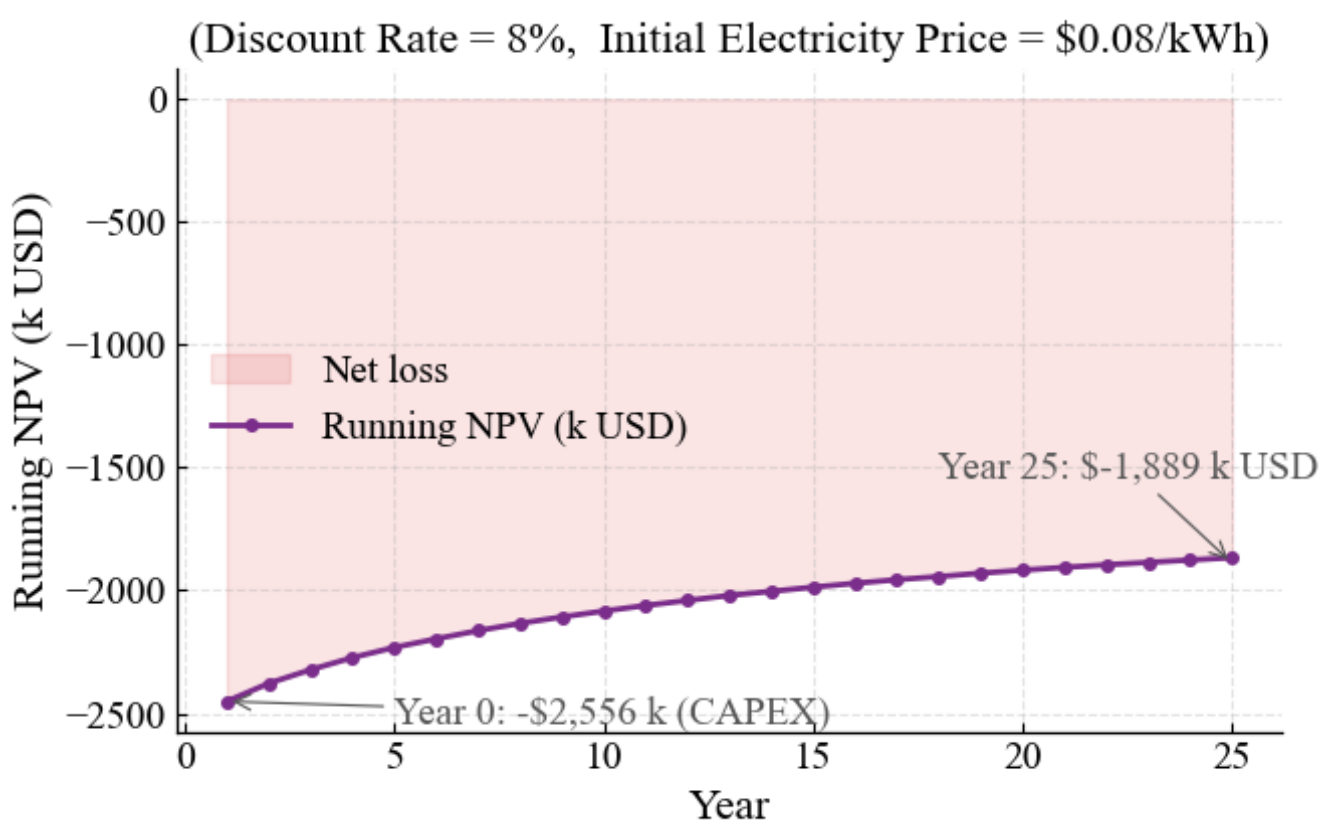

Figure 17. NPV Accumulation over 25 years

## REFERENCES

[1] "Agency, International Energy, Electricity 2024. 2024, IEA: Paris." [Online]. Available: https://www.iea.org/reports/electricity-2024

[2] "Administration, U.S. Energy Information. Changes in the generation mix help reduce U.S. electric power sector emissions. 2023." [Online]. Available: https://www.eia.gov/outlooks/steo/report/BTL/2023/02-genmix/article.php

[3] U.S. Department of Energy, "Pathways to Commercial Liftoff: Next-Generation Geothermal Power," U.S. Department of Energy, Government Report, 2024. [Online]. Available: https://negpa.org/wp-content/uploads/2024/08/LIFTOFF_DOE_NextGen_Geothermal_v14.pdf

[4] S. I. Mohammad *et al.*, "Expedited and dependable geothermal rock characterization and absolute permeability modeling using advanced data-driven techniques," *Geotherm. Energy*, vol. 13, no. 1, p. 45, Dec. 2025, doi: 10.1186/s40517-025-00371-4.

[5] C. Zhou, G. Liu, K. Lei, and S. Liao, "Detailed effects of reservoir permeability distribution differences on enhanced geothermal systems performance," *J. Hydrol.*, vol. 639, p. 131566, Aug. 2024, doi: 10.1016/j.jhydrol.2024.131566.

[6] M. J. Aljubran and R. N. Horne, "Techno-economics of geothermal power in the contiguous United States under baseload and flexible operations," *Renew. Sustain. Energy Rev.*, vol. 211, p. 115322, Apr. 2025, doi: 10.1016/j.rser.2024.115322.

[7] S. Liu and A. Dahi Taleghani, "Closed-loop geothermal systems: Critical review of technologies, performance enhancement, and emerging solutions," *Renew. Sustain. Energy Rev.*, vol. 225, p. 116177, Jan. 2026, doi: 10.1016/j.rser.2025.116177.

[8] M. Zargartalebi, A. Darzi, A. Kazemi, and D. Sinton, "Closed-loop geothermal system is a potential source of low-carbon renewable energy," *Commun. Earth Environ.*, vol. 6, no. 1, p. 812, Oct. 2025, doi: 10.1038/s43247-025-02729-9.

[9] J. Li, H. Wang, T. Wang, H. Li, C. Guo, and D. Yang, "Feasibility analysis of converting abandoned oil and gas wells into geothermal wells and power generation," *Geoenergy Sci. Eng.*, vol. 253, p. 213985, Oct. 2025, doi: 10.1016/j.geoen.2025.213985.

[10] I. Chandra, G. Golla, K. Mccarthy, J. Scherer, H. Chin, and R. Klenner, "Repurposing oil and gas wells to produce electricity using Closed-Loop Geothermal (CLG) technology," May 2023.

[11] M. Raissi, P. Perdikaris, and G. E. Karniadakis, "Physics-informed neural networks: A deep learning framework for solving forward and inverse problems involving nonlinear partial differential equations," *J. Comput. Phys.*, vol. 378, pp. 686–707, Feb. 2019, doi: 10.1016/j.jcp.2018.10.045.

[12] K. Adhikari, M. K. Mudunuru, and K. B. Nakshatrala, "Closed-loop geothermal systems: Modeling and predictions," 2024, *arXiv*. doi: 10.48550/ARXIV.2407.04716.

[13] M. White *et al.*, "Numerical investigation of closed-loop geothermal systems in deep geothermal reservoirs," *Geothermics*, vol. 116, p. 102852, Jan. 2024, doi: 10.1016/j.geothermics.2023.102852.

[14] T. Nassan, H.-C. Turunc, and M. Amro, "Numerical Simulation of Deep Multilateral Closed-Loop Well System for Geothermal Energy Extraction," in *Middle East Oil, Gas and Geosciences Show (MEOS GEO)*, Manama, Bahrain: SPE, Sep. 2025, p. D011S017R005. doi: 10.2118/227167-MS.

[15] H. J. Ramey, "Wellbore Heat Transmission," *J. Pet. Technol.*, vol. 14, no. 04, pp. 427–435, Apr. 1962, doi: 10.2118/96-PA.

[16] M. Habib, A. Habib, and B. Alibrahim, "Applications of physics-informed neural networks in geosciences: From basic seismology to comprehensive environmental studies," *Open Geosci.*, vol. 17, no. 1, p. 20250853, Aug. 2025, doi: 10.1515/geo-2025-0853.

[17] S. Manavi, T. Becker, and E. Fattahi, "Enhanced surrogate modelling of heat conduction problems using physics-informed neural network framework," *Int. Commun. Heat Mass Transf.*, vol. 142, p. 106662, Mar. 2023, doi: 10.1016/j.icheatmasstransfer.2023.106662.

[18] K. Laugksch, P. Rousseau, and R. Laubscher, "A PINN Surrogate Modeling Methodology for Steady-State Integrated Thermofluid Systems Modeling," *Math. Comput. Appl.*, vol. 28, no. 2, p. 52, Mar. 2023, doi: 10.3390/mca28020052.

[19] W. Yuan, Z. Chen, S. E. Grasby, and E. Little, "Closed-loop geothermal energy recovery from deep high enthalpy systems," *Renew. Energy*, vol. 177, pp. 976–991, Nov. 2021, doi: 10.1016/j.renene.2021.06.028.

[20] J. Song, P. Loo, J. Teo, and C. N. Markides, "Thermo-Economic Optimization of Organic Rankine Cycle (ORC) Systems for Geothermal Power Generation: A Comparative Study of System Configurations," *Front. Energy Res.*, vol. 8, p. 6, Feb. 2020, doi: 10.3389/fenrg.2020.00006.

[21] W. Short, D. J. Packey, and T. Holt, "A Manual for the Economic Evaluation of Energy Efficiency and Renewable Energy Technologies," National Renewable Energy Laboratory, NREL/TP-462-5173, 1995.

[22] IEA and NEA, "Projected Costs of Generating Electricity 2020," International Energy Agency, 2020. [Online]. Available: https://www.iea.org/reports/projected-costs-of-generating-electricity-2020

[23] S. Gkousis, K. Braimakis, P. Nimmegeers, S. Karellas, and T. Compernolle, "Multi-objective optimization of medium-enthalpy geothermal Organic Rankine Cycle

plants," *Renew. Sustain. Energy Rev.*, vol. 210, p. 115150, Mar. 2025, doi: 10.1016/j.rser.2024.115150.

[24] K. F. Beckers and H. E. Johnston, "Techno-Economic Performance of Eavor-Loop 2.0," 2022. [Online]. Available: https://api.semanticscholar.org/CorpusID:247165909

[25] H. Wu, J. Xu, J. Wang, and M. Long, "Autoformer: Decomposition Transformers with Auto-Correlation for Long-Term Series Forecasting," *ArXiv CsLG*, 2022.

[26] Orkustofnun (National Energy Authority of Iceland), "Geothermal Well Data and Temperature Profiles Database." 2023. [Online]. Available: https://orkustofnun.is

[27] C. Clauser and E. Huenges, "Thermal Conductivity of Rocks and Minerals," in *AGU Reference Shelf*, T. J. Ahrens, Ed., Washington, D. C.: American Geophysical Union, 2013, pp. 105–126. doi: 10.1029/RF003p0105.

[28] S. W. Scott *et al.*, "Valgarður: a database of the petrophysical, mineralogical, and chemical properties of Icelandic rocks," *Earth Syst. Sci. Data*, vol. 15, no. 3, pp. 1165–1195, 2023, doi: 10.5194/essd-15-1165-2023.

[29] T. L. Bergman, A. S. Lavine, F. P. Incropera, and D. P. DeWitt, *Introduction to Heat Transfer*. Wiley, 2011. [Online]. Available: https://books.google.com/books?id=YBaNaLurTD4C

[30] F. P. Incropera, D. P. DeWitt, T. L. Bergman, and A. S. Lavine, *Fundamentals of Heat and Mass Transfer*, 8th ed. Wiley, 2017.

[31] A. Haghighi, M. R. Pakatchian, M. E. H. Assad, V. N. Duy, and M. Alhuyi Nazari, "A review on geothermal Organic Rankine cycles: modeling and optimization," *J. Therm. Anal. Calorim.*, vol. 144, no. 5, pp. 1799–1814, Jun. 2021, doi: 10.1007/s10973-020-10357-y.

[32] A. S. Laghari, M. W. Chandio, L. Kumar, and M. E. H. Assad, "Thermodynamic Performance Analysis of Geothermal Power Plant Based on Organic Rankine Cycle (ORC) Using Mixture of Pure Working Fluids," *Energy Eng.*, vol. 0, no. 0, pp. 1–10, 2024, doi: 10.32604/ee.2024.051082.

[33] Ormat Nevada Inc. and U.S. Department of Energy, "Final Report: Ormat Organic Rankine Cycle (ORC) Power System Validation Project," Rocky Mountain Oilfield Testing Center (RMOTC), 2007. [Online]. Available: https://geocom.geonardo.com/assets/elearning/7.24.Ormat_report.pdf

[34] R. DiPippo, "ORC Binary Plant Design," in *Proceedings World Geothermal Congress 2020*, International Geothermal Association, 2020.

[35] I. H. Bell, J. Wronski, S. Quoilin, and V. Lemort, "Pure and Pseudo-Pure Fluid Thermophysical Property Evaluation and the Open-Source Thermophysical Property Library CoolProp," *Ind. Eng. Chem. Res.*, vol. 53, no. 6, pp. 2498–2508, 2014, doi: 10.1021/ie4033999.

[36] Meteostat, "Meteostat Python Library." 2024. [Online]. Available: https://dev.meteostat.net/

[37] J. Bull, J. Pound, J. Radulovic, and J. M. Buick, "Low-Temperature ORC Systems: Influence of the Approach Point and Pinch Point Temperature Differences," *Energies*, vol. 18, no. 11, p. 2954, Jun. 2025, doi: 10.3390/en18112954.

[38] J. Sarkar, "A Novel Pinch Point Design Methodology Based Energy and Economic Analyses of Organic Rankine Cycle," *J. Energy Resour. Technol.*, vol. 140, no. 5, p. 052004, May 2018, doi: 10.1115/1.4038963.

[39] T. Akiba, S. Sano, T. Yanase, T. Ohta, and M. Koyama, "Optuna: A next-generation hyperparameter optimization framework," in *Proceedings of the 25th ACM SIGKDD International Conference on Knowledge Discovery & Data Mining*, 2019, pp. 2623–2631.

[40] J. Bergstra, R. Bardenet, Y. Bengio, and B. Kégl, "Algorithms for hyper-parameter optimization," in *Advances in Neural Information Processing Systems (NeurIPS)*, 2011.

[41] Ormat Technologies Inc., "Binary Cycle Power Plants: ORC Technology and Applications." 2019. [Online]. Available: https://geocom.geonardo.com/assets/elearning/7.24.Ormat_report.pdf

[42] *Geothermal Power Plants*. Elsevier, 2016. doi: 10.1016/C2014-0-02885-7.

[43] U.S. Department of Energy, "GeoVision: Harnessing the Heat Beneath Our Feet," U.S. Department of Energy, May 2019. doi: 10.2172/1879171.

[44] International Renewable Energy Agency (IRENA), "Geothermal Power: Technology Brief," IRENA, Abu Dhabi, 2017. [Online]. Available: https://www.irena.org/publications/2017

[45] U. S. E. I. Administration, "Annual Energy Outlook 2025: With projections to 2050," U.S. Energy Information Administration, 2025. [Online]. Available: https://www.eia.gov/outlooks/aeo/